\documentclass{emulateapj} 
\usepackage{color} 
\usepackage[colorlinks,urlcolor=blue,citecolor=blue,linkcolor=blue]{hyper ref}

\usepackage{epsfig}
\usepackage{subfigure}
\usepackage{graphicx}
\usepackage{amsmath}
\usepackage{natbib}
\newcommand{\pa}{\partial}
\newcommand{\mb}{\boldsymbol}
\newcommand{\mc}{\mathcal}

\shorttitle{MHD-PIC Method for CR-Gas Interaction}
\shortauthors{X.-N. Bai et al.}

\begin{document}


\title{Magnetohydrodynamic-Particle-in-Cell Method for
Coupling Cosmic Rays with a Thermal Plasma: Application to Non-relativistic Shocks}


\author{Xue-Ning Bai\altaffilmark{1,3}, 
Damiano Caprioli\altaffilmark{2}, Lorenzo Sironi\altaffilmark{1,4}, Anatoly Spitkovsky\altaffilmark{2}}
\affil{$^1$Institute for Theory and Computation, Harvard-Smithsonian
Center for Astrophysics, 60 Garden St., MS-51, Cambridge, MA 02138}
\affil{$^2$Department of Astrophysical Sciences, Peyton Hall, Princeton
University, Princeton, NJ 08544}
\email{xbai@cfa.harvard.edu}

\altaffiltext{3}{NASA Hubble Fellow}
\altaffiltext{4}{NASA Einstein Fellow}




\begin{abstract}
We formulate a magnetohydrodynamic-particle-in-cell (MHD-PIC) method for 
describing the interaction between collisionless cosmic ray (CR) particles and
a thermal plasma. The thermal plasma is treated as a fluid, obeying equations
of ideal MHD, while CRs are treated as relativistic Lagrangian particles
subject to the Lorentz force. Backreaction from CRs to the gas is included in the
form of momentum and energy feedback. In addition, we include the electromagnetic
feedback due to CR-induced Hall effect that becomes important when the electron-ion
drift velocity of the background plasma induced by CRs approaches the Alfv\'en
velocity. Our method is applicable on scales much larger than the ion inertial length,
bypassing the microscopic scales that must be resolved in conventional PIC methods,
while retaining the full kinetic nature of the CRs. We have implemented and tested
this method in the Athena MHD code, where the overall scheme is second-order
accurate and fully conservative. As a first application, we describe a
numerical experiment to study particle acceleration in non-relativistic shocks.
Using a simplified prescription for ion injection, we reproduce the shock
structure and the CR energy spectra obtained with more self-consistent hybrid-PIC
simulations, but at substantially reduced computational cost. We also show that
the CR-induced Hall effect reduces the growth rate of the Bell instability and
affects the gas dynamics in the vicinity of the shock front. As a step forward,
we are able to capture the transition of particle acceleration from
non-relativistic to relativistic regimes, with momentum spectrum
$f(p)\propto p^{-4}$ connecting smoothly through the transition, as expected from
the theory of Fermi acceleration.
\end{abstract}



\section{Introduction}\label{sec:intro}

Cosmic rays (CRs) are charged particles that possess much higher energy than particles
in the surrounding medium. They are a dynamically important constituent in the Galaxy,
where their energy density is in rough equipartition with gas pressure and magnetic pressure.
They also provide key observational diagnostics for astrophysical
phenomena via non-thermal radiation. The origin and transport of CRs are among the
forefront topics of astrophysics, and involve collisionless kinetic effects with complex
non-linear interactions with magnetic fields and background thermal plasma.

Particle-in-cell (PIC) methods (e.g., \citealp{BirdsallLangdon05}), which iteratively move
particles and update the electromagnetic fields, have been commonly
employed to study the plasma physics of CRs from first principles. Fully kinetic PIC codes
treat both electrons and ions as particles (e.g., \citealp{OSIRIS}, \citealp{VPIC},
\citealp{Spitkovsky05}). While being the
most self-consistent method that rigorously solves for everything from plasma oscillations
to the propagation of light waves, the main disadvantage is that it has to resolve the
microscopic plasma scale: the plasma (electron) skin depth $c/\omega_{pe}\sim5.4\times10^5{\rm cm}\ (n/{\rm cm}^{-3})^{-1/2}$, where
$\omega_{pe}$ is the electron plasma frequency. Moreover, the ion scale is well separated
from the electron scale. Even with artificially reduced ion-to-electron
mass ratio, properly capturing the physics at both scales still demands tremendous
computational cost. 

An alternative approach, called the hybrid-PIC, treats all ions as kinetic particles, while 
electrons are included as a neutralizing, massless and conducting fluid with a prescribed
equation of state (e.g.,
\citealp{Lipatov02,Winske_etal03,Gargate_etal07,Kunz_etal14}). By compromising the
electron-scale physics, the physics at the ion-scale can be well captured with much reduced
computational cost, and this method has been successfully applied to study a wide range of
laboratory, space physics and astrophysical problems.
The hybrid-PIC approach again is limited by the requirement of resolving the characteristic
ion scale, the ion inertial length
$c/\omega_{pi}\sim2.3\times10^7{\rm cm}\ (n/{\rm cm}^{-3})^{-1/2}$,
where $\omega_{pi}$ is the ion plasma frequency and $n$ is the ion number density. This
scale is typically much smaller than the CR gyro radius. The latter scale is particularly
relevant for resonant scattering of CRs, yet it becomes progressively more difficult to 
accommodate using the hybrid-PIC method as the CR gyroradius
becomes much larger than the ion inertial length with increasing CR energy. 

Another approach is to treat CRs as non-thermal test particles, embedded
in a thermal plasma (gas) described by magnetohydrodynamics
(e.g. \citealp{Lehe_etal09,Beresnyak_etal11}).
Being test particles, CRs only passively
respond to the electromagnetic field carried by the gas. On the other hand, CRs are
well known to influence the dynamics of the background gas and magnetic fields, which
in turn affect their propagation (e.g., \citealp{KulsrudPearce69,Skilling75,Parker92}).

In this paper, we consider another approach, which we call the MHD-PIC method.
We treat only CRs as particles using the conventional PIC technique, while the rest of
the thermal plasma (ions and electrons) is treated as gaseous fluid described by
magnetohydrodynamics (MHD). Different from the test particle approach, the CR feedback
to the gas is included following a rigorous formulation that enables the mutual interactions
between the thermal gas and CRs. With CRs treated as particles, kinetic effects of CRs
can be fully captured. With fluid treatment of the thermal plasma, we avoid resolving
microscopic scales, allowing us to accommodate much larger and even macroscopic
scales in our simulations with only modest computational cost\footnote{While the thermal
plasma can be collisionless or weakly collisional, we employ the MHD framework which
implicitly assumes efficient inter-particle ``collisions" via plasma instabilities at microscopic
scales so that the plasma is well isotropized at relatively large scales that we consider.}.


We note that similar methods as we describe in this paper have been proposed and
implemented to study the CR streaming instabilities. \citet{Zachary87} derived the
formulation from a plasma point of view (published in \citealp{ZacharyCohen86}), with
CR feedback reflected in the momentum and energy equations as external Lorentz
force and heating/cooling due to CR charge and current. \citet{LucekBell00} used the
same formulation and discovered in their simulations what was later termed the Bell
instability \citep{Bell04}, which is a non-resonant instability driven by the CR current.
\citet{RevilleBell12} implemented the same method and studied in detail the
filamentation instability of \citet{Bell05}. \citet{RevilleBell13} and \citet{Bell_etal13}
adopted the same MHD formulation, but developed a Vlasov-Fokker-Planck approach
for CRs with approximate closure relations (instead of using particles) to study CR
diffusion, escape and magnetic field amplification in the upstream of collisionless shocks.

In this work, we re-derive the formulation rigorously in the MHD framework, and further
show that there is an additional CR feedback term which modifies the induction and
energy equations. For sufficiently strong CR currents, this term, which we call the
CR-induced Hall term, may appreciably affect the growth rate of the Bell instability 
and become important in the upstream fluid as it approaches the shock front.
We have implemented the MHD-PIC formalism described in this paper into the
state-of-the-art, fully parallelized MHD code Athena \citep{Stone_etal08}. This opens up a
variety of CR-related problems that can be addressed using non-linear computer
simulations, especially the long-term evolution of non-relativistic shocks, to be introduced
in Section \ref{ssec:app}.


\subsection[]{Applications}\label{ssec:app}

One of our main applications is to study the structure, evolution and particle acceleration
in non-relativistic collisionless shocks, which have long been considered as the most
promising sites for producing CRs by means of the first-order Fermi acceleration
mechanism \citep{Fermi49,Bell78a,BlandfordOstriker78}. In particular, diffusive shock
acceleration (DSA) of electrons and nuclei at supernova remnant (SNR) blast waves
is expected to produce power-law energy distributions of accelerated particles extending
over many energy decades \citep[e.g.][]{MorlinoCaprioli12,Ackermann+13}, which is
thought to be responsible for most of the Galactic cosmic rays (CRs), possibly
up to the ``knee" part of the CR spectrum ($\sim10^{15}$eV).

Theoretical investigation of CR acceleration in non-relativistic collisionless shocks has a
long history. Most earlier works either analytically or numerically solve the CR transport
equations (e.g., \citealp{BerezhkoVolk97,Malkov97,Kang_etal02,AmatoBlasi05,ZirakashviliAharonian10,Caprioli12})
or adopt a Monte-Carlo approach (e.g.,
\citealp{Elison_etal90,Elison_etal95,NiemiecOstrowski04,Vladimirov_etal06}). While
these works generally yield results consistent with observations, they rely on
quasi-linear theory for magnetic field amplification with simple prescriptions of CR diffusion
coefficients and injection efficiency, reflecting
the poorly known microphysics of particle injection and turbulence. In particular, the
turbulence that is responsible for scattering CRs should be largely generated by the accelerated
CRs themselves, which in turn affects the shock structure and 
the injection process. The complex interplay among these physical processes calls for
self-consistent numerical simulations.

PIC approach is at the frontier of this research. Simulations have been carried out using 
full PIC codes (e.g., \citealp{AmanoHoshino07,RiquelmeSpitkovsky11,Narayan_etal12,Guo_etal14}),
which mainly address electron acceleration, as well as hybrid-PIC codes (e.g.,
\citealp{Quest88,Giacalone_etal92,Giacalone04,GargateSpitkovsky12,GuoGiacalone13}),
which mainly address ion acceleration. With growing computing power, simulations
with relatively large domains and relatively long duration are becoming more affordable
(e.g., \citealp{CaprioliSpitkovsky13}), allowing the physical processes such as particle
injection, diffusion and acceleration efficiency to be studied in greater detail
\citep{CaprioliSpitkovsky14a,CaprioliSpitkovsky14b,CaprioliSpitkovsky14c}. However,
despite the successes of these state-of-the-art simulations, they still capture only the
initial stages of shock evolution and particle acceleration. Larger spatial scales and longer
time evolution are needed as particles accelerate to higher energies to accommodate the
 interactions of these particles with turbulence. In addition, current hybrid-PIC codes
deal only with non-relativistic particles. Continued acceleration would in reality
lead to transition into the relativistic regime, giving a different power-law slope in 
particle energy spectrum according to the Fermi acceleration theory (from
$f(E)\propto E^{-3/2}$ to $E^{-2}$ in the test particle limit). This transition can not be captured
in hybrid-PIC simulations, and remains to be confirmed. Moreover, since the total energy in
CRs ($\int Ef(E)dE$) would diverge as $E\rightarrow\infty$, we expect the fraction of CR
particles ($\int f(E)dE$) to decrease with time as particles are accelerated to higher energies.
This reduction is important for understanding the long-term evolution of collisionless
shocks, and again is too computationally demanding to be addressed by using the
hybrid-PIC approach.

Without the need to resolve plasma scales, our MHD-PIC approach can substantially
outperform hybrid-PIC codes, allowing us to conduct large simulations following the
long-term evolution of non-relativistic collisionless shocks with self-consistent
particle acceleration. On the other hand, since the gas and CRs are treated separately,
the main disadvantage of the MHD-PIC approach (as well as similar approaches, e.g.,
\citealp{RevilleBell13}) is that injection of CR particles has to be handled with ad hoc
prescriptions. However, this may be remedied by properly designing controlled
experiments to calibrate the injection recipes with hybrid-PIC simulations, as we defer
to future work. In this paper, as the first application of our code to study particle
acceleration in shocks, we adopt a simple and easily controllable injection prescription
to demonstrate the capability of our code to handle highly non-linear regime of CR-gas
interaction. Even with such a simple prescription, we are able to reproduce most of the
essential features of shock structure and particle acceleration as in hybrid-PIC simulations.
Most interestingly, we have performed one simulation with unprecedented long duration
which followed particle acceleration from non-relativistic to relativistic regimes.

In addition, our formalism is well suited to study the interaction between the CRs and the
gas in the ISM, potentially providing microphysical input for studying CR propagation and
transport in the Galaxy \citep{Strong_etal07}. In particular, the non-linear wave-particle
interaction of propagating CRs \citep{KulsrudPearce69} is expected to play an important
role in self-confinement of CRs in the Galaxy. In addition, given the dynamical
importance of (especially low-energy) CRs in the ISM, the back-reaction from CRs
may also affect the cascade of MHD turbulence (e.g., \citealp{LazarianBeresnyak06}),
and in turn their own propagation and transport in the Galaxy. 

This paper is organized as follows. We derive the formulation of the
CR-gas system in Section \ref{sec:eqs}. Implementation of this formulation to the
Athena MHD code is briefly described in Section \ref{sec:imp}, with details and code
tests provided in Appendices \ref{app:imp} and \ref{app:test}. As a special application of
our formulation, we revisit the linear dispersion relation of the Bell instability in
Section \ref{sec:crcdi} and highlight the relevance of the CR-induced Hall term. In
Section \ref{sec:shock}, we describe the setup of shock simulations with a simple
prescription of particle injection. Simulation results presented and analyzed in Sections
\ref{sec:results} and \ref{sec:trans}, highlighting the code capabilities and transition of
particle acceleration from non-relativistic to relativistic regimes.
We summarize and conclude in Section \ref{sec:conclude}.

\section[]{Formulation}\label{sec:eqs}

We consider the interaction between a thermal plasma (gas) and cosmic ray (CR)
particles. The gas is fully ionized, non-relativistic, and is treated as a magnetized
fluid. It consists of ions, which contain all the inertia, and massless electrons,
which mainly conduct current. CR particles are relativistic and collisionless, so
that they are only subject to the Lorentz force. We assume that CRs are particles
with only one sign of charge (positive). While CRs do not interact with
the gas directly, they provide indirect feedback via interactions with
electromagnetic fields. In this section, we formulate the equations for CR-gas
interaction. We rigorously derive the form of CR feedback to the gas in the MHD
framework in Section \ref{ssec:mhdcr}, where CRs are treated as fluid,
and in Section \ref{ssec:crpar}, we discuss the motion of CRs as individual particles.

\subsection[]{MHD Equations with CR Feedback}\label{ssec:mhdcr}

The effect of collisionless CRs on the thermal gas can be well represented by the CR
charge density $n_{\rm CR}$ and current density ${\mb J}_{\rm CR}$ (zeroth and first
moments of momentum distribution, treated as a fluid). They are related by
\begin{equation}
{\mb J}_{\rm CR}=n_{\rm CR}{\mb u}_{\rm CR}\ ,
\end{equation}
where ${\mb u}_{\rm CR}$ can be interpreted as the mean drift velocity of CR particles.

CRs are subject to the Lorentz force, and the force density they experience is
\begin{equation}\label{eq:lorentz}
{\mb F}_{\rm CR}=n_{\rm CR}{\boldsymbol{\mc{E}}}+\frac{1}{c}{\mb J}_{\rm CR}\times{\mb B}\ ,
\end{equation}
where ${\boldsymbol{\mc{E}}}$ and ${\mb B}$ are the electric and magnetic fields.

The total current density in the system comprises of contributions from the gas (ions and
electrons) and the CRs
\begin{equation}
\begin{split}\label{eq:current}
{\mb J}_{\rm tot}=\frac{c}{4\pi}\nabla\times{\mb B}
&={\mb J}_g+{\mb J}_{\rm CR}\\
&=n_i{\mb v}_i+n_e{\mb v}_e+n_{\rm CR}{\mb u}_{\rm CR}\ ,
\end{split}
\end{equation}
where $n_i$ and ${\mb v}_i$ ($n_e$ and ${\mb v}_e$) are the charge density and
mean velocity of the ions (electrons). We are interested in scales
much larger than the Debye length, where charge neutrality applies, 
\begin{equation}\label{eq:charge}
n_i+n_e+n_{\rm CR}=0\ .
\end{equation}
Note that being charge densities, we have $n_i>0$ and $n_e<0$. While we generally
consider CR ions ($n_{\rm CR}>0$), our formulation is equally applicable to
sufficiently energetic CR electrons (e.g., whose energy is higher than energy of thermal
ions), thus $n_{\rm CR}$ can be either positive or negative.

In general, we expect $|n_{\rm CR}|\ll n_i\approx|n_e|$, and this is the regime where
our formulation is applicable: essentially all electrons come from the gas,
thus they are largely thermal and can be treated as a massless fluid, obeying the
force-free condition
\begin{equation}
n_e\bigg({\boldsymbol{\mc{E}}}+\frac{{\mb v}_e}{c}\times{\mb B}\bigg)-\nabla\cdot{\sf P}_e=0\ ,
\end{equation}
where ${\sf P}_e$ is the electron pressure tensor. 

Combining the equations above, we arrive at the generalized Ohm's law in the
presence of CRs:
\begin{equation}
\begin{split}
{\boldsymbol{\mc{E}}}=&-\frac{{\mb v}_g}{c}\times{\mb B}+\frac{1}{|n_e|c}{\mb J}_{\rm tot}\times{\mb B}
-\frac{1}{|n_e|}\nabla\cdot{\sf P}_e\\
&-\frac{n_{\rm CR}}{|n_e|}\frac{({\mb u}_{\rm CR}-{\mb v}_g)}{c}\times{\mb B}\ ,
\end{split}
\end{equation}
where ${\mb v}_g\equiv{\mb v}_i$ is the bulk velocity of the gas (ions).
In the above, the first term is the normal inductive term in MHD, the second
corresponds to the usual Hall term, and the third is the gradient of the electron
pressure.  The last term is a novel term due to the presence of CRs, which we call
the CR-induced Hall term (hereafter, CR-Hall term). It describes the relative drift
between CRs and the bulk of the plasma.

Both the Hall term and the electron pressure gradient term typically
become important only on scales smaller than the ion inertial length $c/\omega_{pi}$,
where $\omega_{pi}=(4\pi n_i^2/\rho)^{1/2}$ is the ion plasma frequency, and $\rho$
is the gas mass density. To see this, we compare the Hall
term with the induction term, and the Hall term becomes important
when ${\mb J}_{\rm tot}/n_e$ is comparable or larger than the Alfv\'en velocity
$v_A=B/\sqrt{4\pi\rho}$. 
Replacing $J_{\rm tot}$ by $cB/L_H$, we see that the Hall term dominates on scales
\begin{equation}\label{eq:LH}
L_H\lesssim\frac{cB}{|n_e|v_A}\approx\frac{c}{\omega_{pi}}\ ,
\end{equation}
where we have used $|n_e|\approx n_i\gg|n_{\rm CR}|$. Similarly, assuming that the
electron pressure is comparable to the gas pressure (which again generally requires
$|n_e|\approx n_i$),
one can further replace $\nabla\cdot{\sf P}_e$ by $\rho c_s^2/L_P$, where $c_s$ is
the sound speed. It is straightforward to show that the electron pressure gradient
term dominates the inductive term when
\begin{equation}\label{eq:LP}
L_P\lesssim\bigg(\frac{c_s^2}{v_A^2}\bigg)\bigg(\frac{c}{\omega_{pi}}\bigg)\ .
\end{equation}
For typical plasmas where thermal and magnetic energies are in rough
equipartition, we find $L_P\sim L_H\sim(c/\omega_{pi})$.

Under the MHD framework, we are interested in scales much larger than
$c/\omega_{pi}$, thus we ignore the conventional Hall and electron
pressure gradient terms. We note that these two terms are explicitly
included in the formulation for hybrid-PIC codes, which aim at resolving the
physics on the scale of ion inertial length. In these codes, some
prescription of the electron pressure tensor (i.e., electron equation of
state) is needed as a closure relation to account for the unresolved electron
microphysics. At much larger scales, however, MHD assumption is expected to
hold (as a closure relation), and we arrive at the ideal MHD Ohm's law with
the corrections due to the CR-Hall term
\begin{equation}\label{eq:emfcr0}
{\boldsymbol{\mc{E}}}=-\frac{{\mb v}_g}{c}\times{\mb B}
-\frac{n_{\rm CR}}{|n_e|}\frac{({\mb u}_{\rm CR}-{\mb v}_g)}{c}\times{\mb B}\ .
\end{equation}
Note that unlike the conventional Hall term, the CR-Hall term does not provide
a typical lengthscale above which it can be safely neglected, and should be
always kept in the formulation.

More insight into the importance of CR-Hall term can be gained by rearranging
the above equation in two ways. First, we define
\begin{equation}
R\equiv\frac{n_{\rm CR}}{|n_e|}=\frac{n_{\rm CR}}{n_i+n_{\rm CR}}\ ,\label{eq:defR}
\end{equation}
the ratio of the CR charge density to the total charge density (of the electrons).
Note that our formulation requires that $R\ll1$. Equation (\ref{eq:emfcr0}) is then
equivalent to
\begin{equation}\label{eq:emfcr1}
{\boldsymbol{\mc{E}}}=-[(1-R){\mb v}_g+R{\mb u}_{\rm CR}]\times{\mb B}/c\ .
\end{equation}
This equation has a clear physical meaning, namely, the advection velocity
of the magnetic field is the charge-density-weighted mean velocity of the
composite gas-CR system. We can also see that the CR-Hall term becomes
dynamically important when the following quantity,
\begin{equation}
\Lambda\equiv\frac{R({\mb u}_{\rm CR}-{\mb v}_g)}{v_A}\ ,\label{eq:lambda}
\end{equation}
approaches unity. While we expect $R\ll1$, CRs may
achieve highly super-Alfv\'enic drift velocities relative to the gas so that
$\Lambda\gtrsim1$. Physically, from Equations
(\ref{eq:current}) and (\ref{eq:charge}), this requires the electron-ion drift
velocity in the thermal gas to be on the order of $v_A$ or larger. Note that
the super-Alfv\'enic streaming between background electrons and ions may lead
to plasma streaming instabilities at the electron scale, e.g., Weibel-like
filamentation was found at about 10 electron skin depths in full-PIC simulations by 
\cite{RiquelmeSpitkovsky09}. These effects are not captured in either MHD or
hybrid-PIC methods. While our formulation remains formally valid for
$\Lambda>1$, one may need to bear in mind potential caveats of unresolved
electron-scale effects.

Second, we note that the CR-Hall term closely resembles the
Lorentz force experienced by CRs. To proceed, we define
\begin{equation}\label{eq:emf0}
{\boldsymbol{\mc{E}}}_0\equiv-\frac{{\mb v}_g}{c}\times{\mb B}\ ,
\end{equation}
as the electric field in ideal MHD without CRs. Substituting Equation
(\ref{eq:emfcr0}) into Equation (\ref{eq:lorentz}), we can re-express
the overall Lorentz force felt by CRs as
\begin{equation}\label{eq:lorentz2}
{\mb F}_{\rm CR}=(1-R)\bigg(n_{\rm CR}{\boldsymbol{\mc{E}}}_0
+\frac{1}{c}{\mb J}_{\rm CR}\times{\mb B}\bigg)\ .
\end{equation}
This is an explicit manifestation of the effect of electromagnetic
feedback: the overall Lorentz force experienced by CRs as a fluid is
reduced by a fraction $R$ compared with the case without the CR-Hall term.
Plugging Equation (\ref{eq:lorentz2}) back to Equation (\ref{eq:emfcr0}),
the electric field can be rewritten in a particularly simple form
\begin{equation}\label{eq:emfcr2}
{\boldsymbol{\mc{E}}}={\boldsymbol{\mc{E}}}_0-\frac{{\mb F}_{\rm CR}}{n_g}\ ,
\end{equation}
where $n_g\equiv n_i$ is the bulk charge density of the (ion) fluid.

So far we have focused on the electromagnetic feedback from the CRs, which
determines the evolution of the magnetic field via the induction equation, 
\begin{equation}
\frac{\pa{\mb B}}{\pa t}=-c\nabla\times{\boldsymbol{\mc{E}}}\ .\label{eq:induction}
\end{equation}
Below we discuss the rest of the MHD equations for the gas.

CRs streaming in the gas also exchange momentum with the gas. Due to
charge-neutrality, the gas (ions and electrons) possesses the opposite charge
density to CRs. Correspondingly, the gas experiences exactly the opposite
electric force density as CRs in any local volume of the system, namely,
$-n_{\rm CR}{\boldsymbol{\mc{E}}}$. In addition, as in Equation (\ref{eq:current}), the total
current density in the system has contributions from both CRs, 
${\mb J}_{\rm CR}$, and the gas, ${\mb J}_g$. The magnetic force experienced by
the gas is ${\mb J}_g\times{\mb B}/c$, while the ideal MHD equations adopt
$(\nabla\times{\mb B})\times{\mb B}/4\pi={\mb J}_{\rm tot}\times{\mb B}/c$.
Therefore, we need to compensate for the difference given by
$-{\mb J}_{\rm CR}\times{\mb B}/c$, which is exactly the opposite of the magnetic
force experienced by CRs. Combining these two facts, we see that the momentum
feedback from CRs on the gas is simply to ``kick" the gas with the opposite of
the Lorentz force ${\mb F}_{\rm CR}$ experienced by CRs. This is an explicit
manifestation of Newton's third law which guarantees total momentum conservation.

The energy feedback from CRs can be understood as the work done by the
external force (CR momentum feedback) per unit time, namely,
${\mb v}_g\cdot(-{\mb F}_{\rm CR})$. Using previous equations (\ref{eq:lorentz}),
(\ref{eq:emfcr1}), (\ref{eq:emf0}), and (\ref{eq:lorentz2}), we can easily derive
\begin{equation}
\begin{split}
{\mb v}_g\cdot{\mb F}_{\rm CR}
&=(1-R){\mb v}_g\cdot(n_{\rm CR}{\boldsymbol{\mc{E}}}_0+{\mb J}_{\rm CR}\times{\mb B}/c)\\
&=(1-R){\mb J}_{\rm CR}\cdot{\boldsymbol{\mc{E}}}_0={\mb J}_{\rm CR}\cdot{\boldsymbol{\mc{E}}}\\
&={\mb u}_{\rm CR}\cdot{\mb F}_{\rm CR}\ .\label{eq:eng}
\end{split}
\end{equation}
We note that ${\mb u}_{\rm CR}\cdot{\mb F}_{\rm CR}$ can be understood as the
rate of energy gain by CRs due to the work done by the Lorentz force. The
above equality means that the same amount of energy is lost by the gas, which is
an explicit manifestation of energy conservation of the composite gas and CR system.
Moreover, the equality with ${\mb J}_{\rm CR}\cdot{\boldsymbol{\mc{E}}}$ indicates
that the energy exchange with the gas is achieved via Joule heating (or cooling).

Based on the discussions above, the remaining of the ideal MHD equations for the gas
can be written as follows (in conservative form)
\begin{equation}\label{eq:gascont}
\frac{\pa\rho}{\pa t}+\nabla\cdot(\rho\mb{v}_g)=0\ ,
\end{equation}
\begin{equation}
\begin{split}
\frac{\pa\rho\mb{v}_g}{\pa t}&+\nabla\cdot(\rho{\mb v}_g^T{\mb v}_g
-\frac{{\mb B}^T{\mb B}}{4\pi}+P_g^*{\sf I})\\
=&-(1-R)(n_{\rm CR}{\boldsymbol{\mc{E}}}_0+{\mb J}_{\rm CR}\times{\mb B}/c)
=-{\mb F}_{\rm CR}\ ,
\end{split}
\label{eq:gasmotion}
\end{equation}
\begin{equation}
\begin{split}
\frac{\pa E}{\pa t}&+\nabla\cdot\bigg[(E+P^*){\mb v}_g
-\frac{({\mb B}\cdot{\mb v}_g){\mb B}}{4\pi}
+\frac{c}{4\pi}({\boldsymbol{\mc{E}}}-{\boldsymbol{\mc{E}}}_0)\times{\mb B}\bigg]\\
=&-(1-R){\mb J}_{\rm CR}\cdot{\boldsymbol{\mc{E}}}_0
=-{\mb u}_{\rm CR}\cdot{\mb F}_{\rm CR}\ ,
\end{split}\label{eq:engeq}
\end{equation}
where $P_g^*\equiv P_g+B^2/8\pi$, ${\sf I}$ is the identity tensor, and the total energy
density is defined as
\begin{equation}
E=\frac{P_g}{\gamma-1}+\frac{1}{2}\rho v_g^2+\frac{B^2}{8\pi}\ .\label{eq:gaseng}
\end{equation}
In the above, $\rho$, $P_g$ are gas density and pressure, and $\gamma$ is the
adiabatic index. There is also a source term in the energy flux in the energy equation
(\ref{eq:engeq}), which accounts for the correction to the ideal MHD Poynting flux
($c{\boldsymbol{\mc{E}}}_0\times{\mb B}/4\pi$) due to the CR-Hall term.

Note that the CR feedback source terms are generally proportional to
$n_{\rm CR}\propto R$. The requirement of $R\ll1$ guarantees that as long as
our formulation remains valid, the source terms would never become stiff
to significantly affect the overall evolution of the gas.

To summarize, we have formulated the set of MHD equations appropriate
for studying the interaction between CRs and the thermal plasma. Our
formulation properly describes the physics of the composite system on
scales much larger than the ion inertial length, $c/\omega_{pi}$, and when the
CR charge and current density satisfy $R\ll1$. 
We highlight that the presence of streaming CRs introduces electromagnetic
feedback in the form of a CR-Hall term to the induction equation characterized
by parameter $\Lambda$ (Equation \ref{eq:lambda}). This term is of the same order
as the momentum and energy feedback from CRs, and becomes dynamically
important when $\Lambda$ is not much less than $1$, with the potential
caveat that uncaptured plasma instabilites at unresolved electron scales
may arise when $\Lambda\gtrsim1$.


\subsection[]{Particle Treatment of CRs}\label{ssec:crpar}

We adopted Gaussian units in our formulation in Section \ref{ssec:mhdcr}, which
contains factors of speed of light, $c$, in the expressions. This factor has
no physical significance in non-relativistic MHD and can be absorbed by redefining
the electric field and 
charge/current density as $c{\boldsymbol{\mc{E}}}\rightarrow {\boldsymbol{\mc{E}}}$
and ${\mb J}/c\rightarrow{\mb J}$.
On the other hand, our CR particles can be fully relativistic. Therefore, we artificially
specify the speed of light ${\mathbb C}$ for the CR particles. For consistency with MHD,
the value of ${\mathbb C}$ must be chosen such that it is much larger than any
characteristic MHD velocities.
We still keep the irrelevant factor $c$ so as to retain consistency and to contrast with the
factor ${\mathbb C}$.\footnote{One may continue to use $c$ to specify particle speed of
light. We prefer to introduce a different symbol for conceptual clarity: in the non-relativistic
MHD framework, ${\mathbb C}$ is a freely-variable numerical parameter rather than a
physical ``constant."}

For individual particle $j$, we use ${\mb u}_j$ to denote its velocity. We further define
the vector component of its four-velocity to be
\begin{equation}
{\mb v}_j\equiv\gamma_j{\mb u}_j\ ,\label{eq:4vel}
\end{equation}
where the Lorentz factor
\begin{equation}
\gamma_j=\frac{\sqrt{{\mathbb C}^2+v_j^2}}{\mathbb C}
=\frac{\mathbb C}{\sqrt{{\mathbb C}^2-u_j^2}}\ .
\end{equation}
We define CR particle kinetic energy $E_k$ as the difference between its total
energy and rest mass energy:
\begin{equation}\label{eq:parEk}
E_{k,j}\equiv(\gamma_j-1){\mathbb C}^2=\frac{v_j^2}{\gamma_j+1}\ ,
\end{equation}
which reduces to standard expressions of particle energy in both non-relativistic
and relativistic regimes.

The particle equation of motion reads
\begin{equation}
\frac{d\mb{v}_j}{d t}
=\bigg(\frac{q}{mc}\bigg)_j\bigg(c{\boldsymbol{\mc{E}}}+{\mb u}_j\times{\mb B}\bigg)\ .\label{eq:CRmotion}
\end{equation}
where $(q/mc)_j$ represents particle charge-to-mass ratio. For the electric field,
instead of using its primitive expression (\ref{eq:emfcr0}), we use the alternative
expression from Equation (\ref{eq:emfcr2}). It has the clear advantage that the
correction from the CR-Hall term is proportional to the momentum feedback
(${\mb F}_{\rm CR}$), and hence no extra computation is needed.
Note that here, and essentially in all other occasions, the factor $c$ can be absorbed
into the $(q/mc)$ factor. This factor incorporates the particle intrinsic properties and is
the only physical parameter to distinguish different CR particle species.

\section[]{Implementation}\label{sec:imp}

We have implemented the MHD formulation described in \S
\ref{sec:eqs} together with kinetic CR particles in the Athena MHD code
\citep{Stone_etal08}. Hereafter, we will refer to our gas-particle framework as
the MHD-PIC.

The Athena MHD code uses higher-order Godunov methods with constrained
transport technique to enforce divergence-free condition on the magnetic field. Two
MHD integrators are implemented in Athena, namely the corner transport
upwind (CTU, \citealp{GardinerStone05,GardinerStone08}) and the van-Leer (VL,
\citealp{StoneGardiner09}) integrators. Both integrators are dimensionally unsplit
and second order accurate in time, with the CTU integrator being more accurate
and less diffusive (but more complicated). We always use the CTU integrator in
our MHD-PIC applications, while changing the framework for the VL integrator
would be a trivial extension.

Lagrangian particles have been implemented in the Athena MHD code in the context
of protoplanetary disks \citep{BaiStone10a}. These particles represent solid bodies
interacting with the gas via aerodynamic drag. Backreaction from the particles to the
gas (feedback) are also properly implemented which is an essential ingredient to study
the streaming instability \citep{YoudinGoodman05} as a pathway to planetesimal
formation \citep{Johansen_etal09,BaiStone10b}. The particle integrator and feedback
are combined with the MHD integrator using a predictor-corrector scheme, which
achieves second-order accuracy for the composite system. Taking advantage of the
existing infrastructure for solid particles with feedback,
we briefly describe our implementation of the CR particles with feedback below,
with more details provided in Appendix \ref{app:imp}.

We employ the standard second-order accurate triangular-shaped cloud (TSC)
scheme based on quadratic spline (QS) interpolation \citep{BirdsallLangdon05}
for interpolation of grid quantities to particle locations and for depositing particle
quantities back to the MHD grid. Standard relativistic Boris integrator is used to
push the particles \citep{Boris70}. The particle integrator is combined with the
MHD integrator with feedback added as source terms in a similar predictor-corrector
scheme as in \citet{BaiStone10a} to achieve second-order accuracy. Exact
conservation of momentum and energy in the composite system of gas and CRs
is achieved by construction under the Godunov framework. We have
further implemented sub-cycling for CR particles to improve the code performance.
In practice, we generally choose 5 particle steps per MHD step considering the
balance between accuracy and efficiency.

We have tested all ingredients of our implementation including the Boris integrator,
CR feedback, and particularly the CR-Hall effect with carefully designed test
problems, which are described in Appendix \ref{app:test}.

\section[]{Linear Analysis of the Bell Instability}\label{sec:crcdi}

Before proceeding to applications, we first consider the non-resonant CR streaming
instability of \citet{Bell04}, and point out the relevance of the CR-Hall term that was
previously neglected. This instability occurs when the CR drift velocity exceeds the
local Alfv\'en velocity, where right-handed circularly-polarized modes become unstable
as the Lorentz force exceeds magnetic tension.
This instability has unstable wavelengths much smaller than the Larmor radius of the
CR particles, and its free energy comes from the relative streaming of gas and CRs.

The Bell instability is considered to play an important role in the upstream of
high Mach number non-relativistic collisionless shocks, such as SNR shocks. 
Accelerated CRs drift with respect to the upstream fluid at velocity $U_s$, which
is of the order the shock velocity or higher and is much larger than $v_A$.
Since the typical growth time is much shorter than the advection time in the CR-induced
shock precursor, the instability typically enters its strongly non-linear stage;
the preshock magnetic field may be effectively amplified, and the self-generated
magnetic turbulence feeds back on CRs, enhancing their diffusion.
Hybrid-PIC simulations have recently shown that such a non-linear interplay between
CRs and magnetic fields shapes the large-scale structure of the shock, and promotes
rapid acceleration of energetic particles
\citep{CaprioliSpitkovsky14a,CaprioliSpitkovsky14b,CaprioliSpitkovsky14c}.

We note that the CR-Hall term was not included either in the original MHD derivation of
the instability \citep{Bell04}, or in the kinetic approach by \cite{AmatoBlasi09}. Here we
re-derive the dispersion relation that properly includes the CR-Hall term. Suppose the
background gas is uniform with density $\rho_0$ and zero velocity ${\mb v}_g=0$
(upstream reference frame). Let ${\mb B}_0$ be the background magnetic field, and
${\mb J}_{\rm CR}=n_{\rm CR}{\mb U}_s$ be the external current density provided by
the CR particles. Both of them are along the $\hat{x}$ direction. Now we consider
incompressible perturbations in the gas, assuming all physical quantities to evolve as
$\exp{{\rm i}(kx-\omega t)}$ (note that the wave vector ${\mb k}$ is also along the $\hat{x}$
direction) and use ${\mb u}$, ${\mb b}$ to denote first-order perturbations of gas velocity
and magnetic field. For incompressible modes, both quantities are in the transverse
direction. From the momentum equation (\ref{eq:gasmotion}) (or more explicitly see
Equation of (1) of \citealp{Bell04}), we have
\begin{equation}
\begin{split}
-{\rm i}\omega\rho_0&({\mb u}\times{\mb k})=
+{\rm i}({\mb k}\cdot{\mb B}_0)({\mb b}\times{\mb k})\\
&-(1-R)[({\mb J}_{\rm CR}\times{\mb b})
-n_{\rm CR}({\mb u}\times{\mb B}_0)]\times{\mb k}\ ,\\
\end{split}
\end{equation}
where for conciseness we have ignored factors containing $c$ and $4\pi$, which
will be irrelevant to the results. Also note the factor $(1-R)$ that comes
from Equation (\ref{eq:lorentz2}).
The perturbation equation for the induction equation (\ref{eq:induction}) reads
\begin{equation}
-\omega{\mb b}=(1-R)({\mb k}\cdot{\mb B}_0){\mb u}
-R({\mb k}\cdot{\mb U}_s){\mb b}\ .
\end{equation}
The above equation gives the relation between ${\mb b}$ and ${\mb u}$.

Combining the two equations, and noting that ${\mb k}$, ${\mb B}_0$ and
${\mb J}_{\rm CR}$ are parallel, we find after some algebra
\begin{equation}
\begin{split}\label{eq:bell1}
&{\rm i}[\omega^2-\omega kRU_s-(1-R)k^2v_A^2]({\mb b}\times\hat{x})\\
+&(1-R)\frac{kJ_{\rm CR}v_A^2}{B_0}\bigg(1-\frac{\omega}{kU_s}\bigg){\mb b}=0\ .
\end{split}
\end{equation}
Note that all terms containing $R$ are due to the CR-Hall term,
and the most significant modification lies in the term $\omega kRU_s$,
which leads to a non-negligible shift in wave frequency. Our $\Lambda$-parameter
defined in Equation (\ref{eq:lambda}) corresponds to $\Lambda=RU_s/v_A$. 

We provide the full derivation of the dispersion relation in Appendix \ref{app:bell}.
In the following, we consider the limit $R\ll1$ with no constraints on $\Lambda$,
and just quote the results from the full derivation (see Equation \ref{eq:belldisp2})
\begin{equation}
(\tilde{\omega}\pm\epsilon k_0v_A)^2
=f\bigg(k\pm \frac{1}{f}k_0\bigg)^2v_A^2-(k_0v_A)^2\bigg(\frac{1}{f}-\epsilon^2\bigg)\ ,\label{eq:belldisps}
\end{equation}
where the two $\pm$ signs must be taken to be the same, and we have defined
\begin{equation}
\tilde{\omega}\equiv\omega-(\Lambda/2)v_Ak\ ,\qquad
k_0\equiv\frac{J_{\rm CR}}{2B_0}\ ,
\end{equation}
\begin{equation}
f\equiv1+(\Lambda/2)^2\ ,\qquad
\epsilon\equiv\frac{v_A}{U_s}\ .
\end{equation}
Note that for highly super-Alfv\'enic shocks, $\epsilon\ll1$. When the CR-Hall term
is not included or for $\Lambda\ll1$, we have $f=1$, and the dispersion relation
(\ref{eq:belldisps}) reduces to Equation (7) of \citet{Bell04} (regime II,
since we treat CRs as a fixed external current). In the general case, the system
becomes unstable when $\epsilon^2<1/f$. The most unstable mode corresponds to
$k_m=\pm k_0/f$, with fastest growth rate of the order $k_0v_A/\sqrt{f}$.

Now we discuss the role of the CR-Hall term, setting $\epsilon=0$, valid for
highly super-Alfv\'enic shocks. Our linear analysis of the Bell instability leads to qualitative
changes in two aspects.
First, the most unstable wavenumber $k_m$ is reduced by a factor of about $f$. Without
the CR-Hall term, we have $k_m=k_0\propto J_{\rm CR}\propto\Lambda$, which
increases with $J_{\rm CR}$ without bound. 
Including this term, and considering the limit
$\Lambda\gg1$ (hence $f\sim(RU_s/2v_A)^2$) we have
\begin{equation}
k_m=k_0/f\sim\frac{2J_{\rm CR}v_A^2}{B_0R^2U_s^2}
\sim\bigg(\frac{c}{\omega_{pi}}\bigg)^{-1}\frac{2}{\Lambda}\ .
\end{equation}
Matching the two limits of $\Lambda\ll1$ and $\Lambda\gg1$, we see that as
$J_{\rm CR}$ (hence $\Lambda$) increases from zero,
$k_m\propto J_{\rm CR}$, and the most unstable wavelength decreases until it
reaches the scale of ion inertial length, at about $\Lambda\sim2$. Further increasing
$J_{\rm CR}$ would increase the most unstable wavelength. Therefore, the most
unstable mode never falls below the scale of $c/\omega_{pi}$. This result may have
important consequences for the resonant scattering of CRs in the vicinity of the
shock front.
Second, the fastest growth rate $\sim k_0v_A/\sqrt{f}$, is reduced by a factor of
about $\sqrt{f}$. Without the CR-Hall term, the fastest growth rate increases
with $J_{\rm CR}$ without bound since $k_0\propto J_{\rm CR}$. Including this term,
and again in the limit $\Lambda\gg1$, we have
\begin{equation}
\omega_{\rm max}\sim\frac{J_{\rm CR}v_A^2}{B_0RU_s}
\sim\Omega_c\ .
\end{equation}
Therefore, the maximum growth rate saturates at $\Omega_c\equiv qB_0/m_ic$,
the ion cyclotron frequency of the background plasma.

We note that using full kinetic PIC simulations, \citet{RiquelmeSpitkovsky09} showed
that when $\Lambda\gtrsim1$, the Bell instability can be overtaken by a Weibel-like
filamentation instability, found earlier by \citet{Niemiec_etal08}. Given the fact that the
growth of the Bell mode saturates at $\Omega_c$ and migrates to larger scales at
$\Lambda\gg1$, it is not surprising that the Bell mode is overtaken by the filamentation
mode, which may not be captured in our MHD-PIC formulation.

We will show in our shock simulations that $\Lambda$ can approach order unity in
the vicinity of the shock front. While this already corresponds to the non-linear stage
of the Bell instability, our derivations in the linear regime still provide a useful quantitative
assessment on the importance of the CR-Hall effect on the gas dynamics in this region.

\section[]{Particle Acceleration in Collisionless Shocks: Simulation Setup}\label{sec:shock}

As introduced in Section \ref{ssec:app}, we conduct a proof-of-concept study of
the evolution and particle acceleration in non-relativistic collisionless shocks to
demonstrate our code performance, where the non-linear development of the Bell
instability in the shock upstream plays a vital role
(e.g., \citealp{Bell_etal13,CaprioliSpitkovsky14b}).

The basic setup of the shock problem involves colliding a highly super-Alfv\'enic
flow into a conducting and static wall (left boundary in the Figures). The shock
forms instantaneously,
propagating away from the wall. After the collision, the downstream gas has zero
longitudinal velocity, hence the simulation takes place in the downstream frame.
Let $\hat{x}$ denote the direction of the shock, and the conducting wall is located
at $x=0$. The colliding gas (upstream) moves from the right ($x>0$) toward the wall,
with uniform density $\rho_0$, pressure $p_0$ and velocity $-v_0\hat{x}$.
We consider parallel shocks with magnetic field along the $\hat{x}$ direction, whose
strength is $B_0$. For convenience,
we define the shock Alfv\'enic Mach number to be $M_A\equiv v_0/v_{A0}$, where
$v_{A0}=B_0/\sqrt{\rho_0}$ is the background Alfv\'en velocity\footnote{More strictly,
shock Mach number is defined in the shock frame, our definition simply follows the
convention of \cite{CaprioliSpitkovsky13}.}. The angle between
the background field and shock normal ($\hat{x}$) is denoted by $\theta$ which,
together with $M_A$, are the main shock parameters (the sonic Mach number is
comparable to $M_A$).

For the purposes of this paper, we only consider high Mach number, parallel
shocks with $M_A=30$. For highly super-sonic and
super-Alfv\'enic shock, we have $\rho_0v_0^2\gg P_0+B_0^2/2$,
which leads to shock jump conditions (expressed in the downstream frame)
under ideal MHD
\begin{equation}
\begin{split}
&\rho_1=r\rho_0\ ,\qquad\qquad P_1=\frac{r}{r-1}\rho_0v_0^2\ ,\\
&B_{x1}=B_{x0}\ ,\qquad\quad  
v_{sh}=v_0/(r-1)\ .
\end{split}\label{eq:rh}
\end{equation}
where $r\equiv(\gamma+1)/(\gamma-1)$ is the shock compression ratio,
$v_{sh}$ is the velocity of the shock in the simulation (downstream) frame.
For adiabatic index $\gamma=5/3$, we
have $r=4$ for strong shock as usual. We use subscripts ``$0$" and ``$1$" to
denote upstream and downstream quantities measured in the simulation (downstream)
frame, hence $v_1=0$ by definition. The shock
converts the kinetic energy in the upstream gas into thermal energy.
The gas temperature can be defined as $p/\rho$, hence $T_0=P_0/\rho_0$,
while $T_1=v_0^2/(r-1)$ is independent of $T_0$ for strong shocks. We also
define the shock kinetic energy $E_{sh}\equiv v_0^2/2$, which characterizes
the energy of downstream thermal particles (per unit mass) and will be used
as a scale to normalize the energy of CR particles.

The natural units for length, time and velocity are the ion inertial length
$c/\omega_{pi}=(q/mc)^{-1}\rho_0^{-1/2}$, ion cyclotron frequency
$\Omega_{c0}=(q/mc)B_0$, and the Alfv\'en velocity $v_{A0}=B_0\rho_0^{-1/2}$.
They are all set to $1$ in code units, with $\rho_0=B_0=(q/mc)=1$. We also
set $T_0=1$ (or $P_0=1$), which is irrelevant as long as $T_0\ll v_0^2$.
We are mainly interested in highly super-Alfv\'enic shocks so that $v_0\gg v_{A0}$
(we choose $v_0=30$). This means that in the downstream frame, the gyro-radius of
a thermal particle
around background field is much larger than $c/\omega_{pi}$, by a factor of
$\sim M_A$. Initially, we focus on particle acceleration
in the non-relativistic regime, and set the particle speed of light to 
${\mathbb C}=10^4\gg v_0$ (so that particles are not able to achieve relativistic
velocities within the duration of simulations). In addition, we also run
one simulation with much reduced speed of light ${\mathbb C}=10\sqrt{2}v_0=424.3$
(or equivalently larger physical shock velocity). In this case, the energy for
non-relativistic to relativistic transition $E_{k,{\rm trans}}\sim{\mathbb C}^2/2$ is
about $200E_{sh}$, which is optimal for studying particle acceleration trasitioning to
the relativistic regime.

Being a higher-order Godunov code, the MHD integrator in Athena works very
well for shock capturing. With uniform and homogeneous upstream medium, the
shock is captured with only 2-3 grid cells using the CTU integrator with third-order
reconstruction. The microphysics underlying the sharp transition in the MHD
shock is the efficient particle thermalization process, mediated by electromagnetic
turbulence as a result of plasma instabilities on microscopic scales
($\lesssim c/\omega_{pi}$). While such plasma micro-instabilities are not captured in
the MHD framework, we expect the scattering of CR particles to be dominated
by turbulence on larger scales in highly super-Alfv\'enic turbulence, which is
captured in the MHD-PIC approach when streaming instabilities such as the Bell
instability are fully developed.

\begin{table}
\caption{List of main shock simulation runs}\label{tab:shock}
\begin{center}
\begin{tabular}{c|cccc}\hline\hline
 Run & Domain size& resolution &  $\eta$ &  ${\mathbb C}$ \\
   & $L_x\times L_y$ ($c/\omega_{pi}$) & ($c/\omega_{pi}$ per cell) &   &  \\\hline

R1 & $(1.2\times10^5)\times3000$ & $12$ & $10^{-3}$ & $10^4$ \\
R2 & $(1.2\times10^5)\times3000$ & $12$ & $2\times10^{-3}$ & $10^4$ \\
R4 & $(1.2\times10^5)\times3000$ & $12$ & $4\times10^{-3}$ & $10^4$ \\
R2-hr & $(1.2\times10^5)\times3000$ & $6$ & $2\times10^{-3}$ & $10^4$ \\
R2-REL & $(3.89\times10^5)\times4800$ & $12$ & $2\times10^{-3}$ & $424.3$ \\
\hline\hline
\end{tabular}
\end{center}

Note: all simulations correspond to parallel shocks with Alfv\'enic Mach
number $M_A=30$.

\end{table}

Here we describe our simple prescription for particle injection. In every time
step, we keep track of the location of the shock front $x_s$ by computing the
transversely averaged profile of $v_x$ along $\hat{x}$. This 1D profile is further
smoothed with a Gaussian kernel whose width is 4 grid cells. The shock
front $x_s$ is identified as the location where $v_x$ is reduced by about $40\%$
with respect to the upstream value.
With $x_s$ identified, the amount of mass traversed by the shock can also be
obtained. We inject a spatially uniform, isotropic distribution of CR particles at the
shock front in the co-moving frame with the shock, whose energy is set to be
$10E_{sh}$ in this frame. The amount (mass fraction) of CR particles we inject is
a small and fixed fraction $\eta$ of the amount of mass swept by the shock. Note
that $E_k=10E_{sh}$ is approximately the threshold energy for a particle to be
considered as non-thermal and to participate in the DSA process
\citep{CaprioliSpitkovsky14a,Caprioli_etal15}. In brief, $\eta$ is our only
parameter for particle injection, which encapsulates our ignorance of the
physics of particle injection.

Physically, the injected particles originate from the gas that is newly processed
by the shock. While we do not capture the injection physics with the simple
prescription, we do compensate for the mass, momentum and energy of
injected particles by subtracting them off from the gas at each timestep over
certain width at the downstream side of the shock\footnote{We further define the
location of the tail end of the shock $x_t$ based on the 1D smoothed pressure
profile (where $P$ achieves $2P_1/3$). Compensation is performed largely
between $x_t$ and $x_s$.}, so that the total mass, momentum and energy in
the full system are conserved.
For typical injection fraction $\eta=2\times10^{-3}$ (see below), because the
energy of injected particles only amounts to a very small fraction of the shock
energy ($\sim3\%$), influences to shock dynamics from this procedure is
hardly noticeable. Also, the simulation outcome is very insensitive to the
detailed implementation of the compensation scheme.

\begin{figure*}
    \includegraphics[width=180mm]{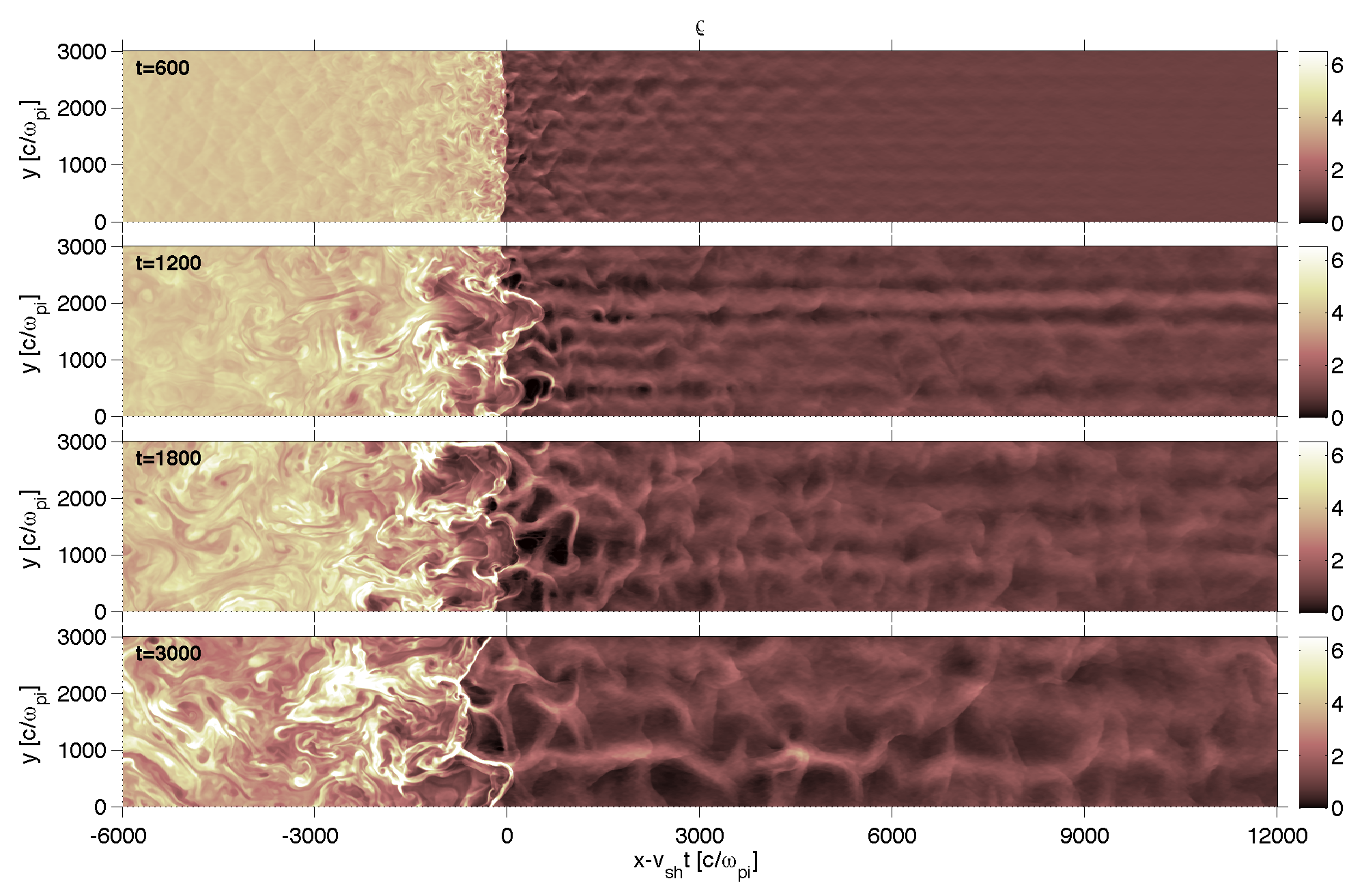}
  \caption{Snapshots of gas density at four different times (as indicated) from
  our fiducial run R2 which show the evolution of the shock. Only a small
  fraction of the simulation box is shown which is adjusted to cover the vicinity
  of the expected shock location at $x=v_{sh}t=v_0t/3$.}\label{fig:evolve}
\end{figure*}

In the simulations, before the Bell instability is fully developed, most of the
injected particles travel freely into the shock upstream. These streaming particles,
on the one hand, source the initial growth of the Bell instability, while on the
other hand, they are not effectively scattered since they only experience unperturbed
or very weakly perturbed magnetic field. Particles injected later experience fully
developed turbulence from the Bell instability and are efficiently scattered to enter
DSA. As a result, there is a transient flow of unperturbed CR particles
traveling through the upstream medium (with approximately constant $j_{\rm CR}$).
Since the growth rate of the Bell instability is proportional to $j_{\rm CR}$, this
transient CR flow would lead to growth of the Bell instability by a finite amplitude,
as a result of the unrealistic initial condition. To remedy this artifact, we record the
injection time $t_i$ for every CR particle. All particles with $t_i<480\Omega_c^{-1}$
(which is more than sufficient for the Bell instability to be fully developed near the
shock front) are removed at time $t=2t_i$. Following this procedure, we expect
our results to be largely unaffected by initial conditions at $t>960\Omega_c^{-1}$.

We perform 2D shock simulations with fiducial domain size
$L_x=1.2\times10^5(c/\omega_{pi})$ and $L_y=3000(c/\omega_{pi})$. We use
conducting boundary condition at $x=0$, and inflow boundary condition at
$x=L_x$, where all quantities are fixed at the initial value. With $M_A=30$,
hence $v_{sh}=10$, we run the simulations up to time $t=3000\Omega_{c0}^{-1}$
so that at the end of the simulation, the gas initially at the right boundary just
reaches the shock.
We choose fiducial numerical resolution of 12$c/\omega_{pi}$ per cell, which
will be shown to well capture the most unstable modes of the Bell instability.
This is also much smaller than the Larmor radius of the lowest energy CR particles
orbiting background field (for $E_k=10E_{sh}$ particles, the Larmor radius
$R_L\sim100c/\omega_{pi}$). We also perform a resolution study with up to four
times higher and two times lower resolution (see Section \ref{ssec:res}), while we
mainly discuss the run with twice resolution.
Note that in hybrid-PIC simulations, the typical numerical resolution is 2 cells
per $c/\omega_{pi}$ in order to properly capture the essential microphysics
(e.g., \citealp{GargateSpitkovsky12,CaprioliSpitkovsky13}). With the MHD-PIC
approach, the gain in efficiency is tremendous.\footnote{In addition,
the particle timestep in our MHD-PIC simulations is mainly constrained
by resolving the Larmor time and grid-crossing time, and with
$\Upsilon=0.3$ (see Appendix \ref{app:dt}), the typical particle
timestep $\Delta t>0.006\Omega_{c0}^{-1}$ is more than 6 times
larger than used in hybrid-PIC.}
For the simulation with reduced speed of light ${\mathbb C} =10\sqrt{2}v_0=424.3$,
we enlarge the domain size in both dimensions to
$L_x=3.888\times10^5(c/\omega_{pi})$, $L_y=4.8\times10^3(c/\omega_{pi})$, and
run the simulation to $t=11520\Omega_{c0}^{-1}$. 

We choose $\eta=2\times10^{-3}$ as the standard CR mass fraction in our
simulations. This is motivated by the numbers obtained from hybrid-PIC
simulations \citep{CaprioliSpitkovsky14a}. In these simulations, it was found
that in a parallel shock, about $\xi\approx15\%$ of the shock kinetic energy is
converted to non-thermal particles. With $\eta=2\times10^{-3}$, the energy
contained in our initially injected particles is about $3\%$ of the shock energy.
This is sufficiently small so as not to affect the shock structure, leaving enough
room for these particles to further gain energy from the Fermi acceleration
process.
Moreover, the most unstable wavelength of
the Bell instability in the shock vicinity can be estimated to be
$\lambda_m\sim3\pi/(\eta M_A)\sim157(c/\omega_{pi})$, which is well resolved
with $\sim13$ cells. Larger $\lambda_m$ is expected further upstream of the shock
due to the reduction of net CR current density. As a parameter study, we also
perform simulations with $\eta=10^{-3}$ and $\eta=4\times10^{-3}$.
For all simulations, the number of particles we inject is equivalent to 4
particles per cell at background density $\rho_0$, or 16 particles per cell on
average in the downstream. Note that despite the reduced spatial resolution, the
number density of CR particles in physical units is comparable to that in typical
hybrid-PIC simulations (where spatial resolution is much higher but only a
very small fraction of the particles enters Fermi acceleration),
and we have verified that simulation results converge with respect to number of
injected particles.
In the simulations, we allow a maximum of 5 particles steps per MHD
step to speed up the calculation. 

In sum, we mainly discuss 5 simulation runs: three runs with
${\mathbb C}=10^4\gg v_0$ using standard resolution with
$\eta=10^{-3}, 2\times10^{-3}$ and $4\times10^{-3}$,  one high resolution run,
and one run with reduced speed of light ${\mathbb C}=10\sqrt{2}v_0=424.3$.
These runs are summarized in Table \ref{tab:shock}, and each run is assigned a
run name as indicated. We consider run R2 as the fiducial simulation and other
runs as variations. All runs except for run R2-REL are relatively cheap
computationally, with our fiducial run R2 taking about $2700$ CPU hours on
a HP Beowulf cluster with
3.47 GHz Intel Westmere cores, yet the simulation box size and duration are
already much larger than in the state-of-the-art hybrid-PIC simulations.

\begin{figure*}
    \centering
    \includegraphics[width=89mm]{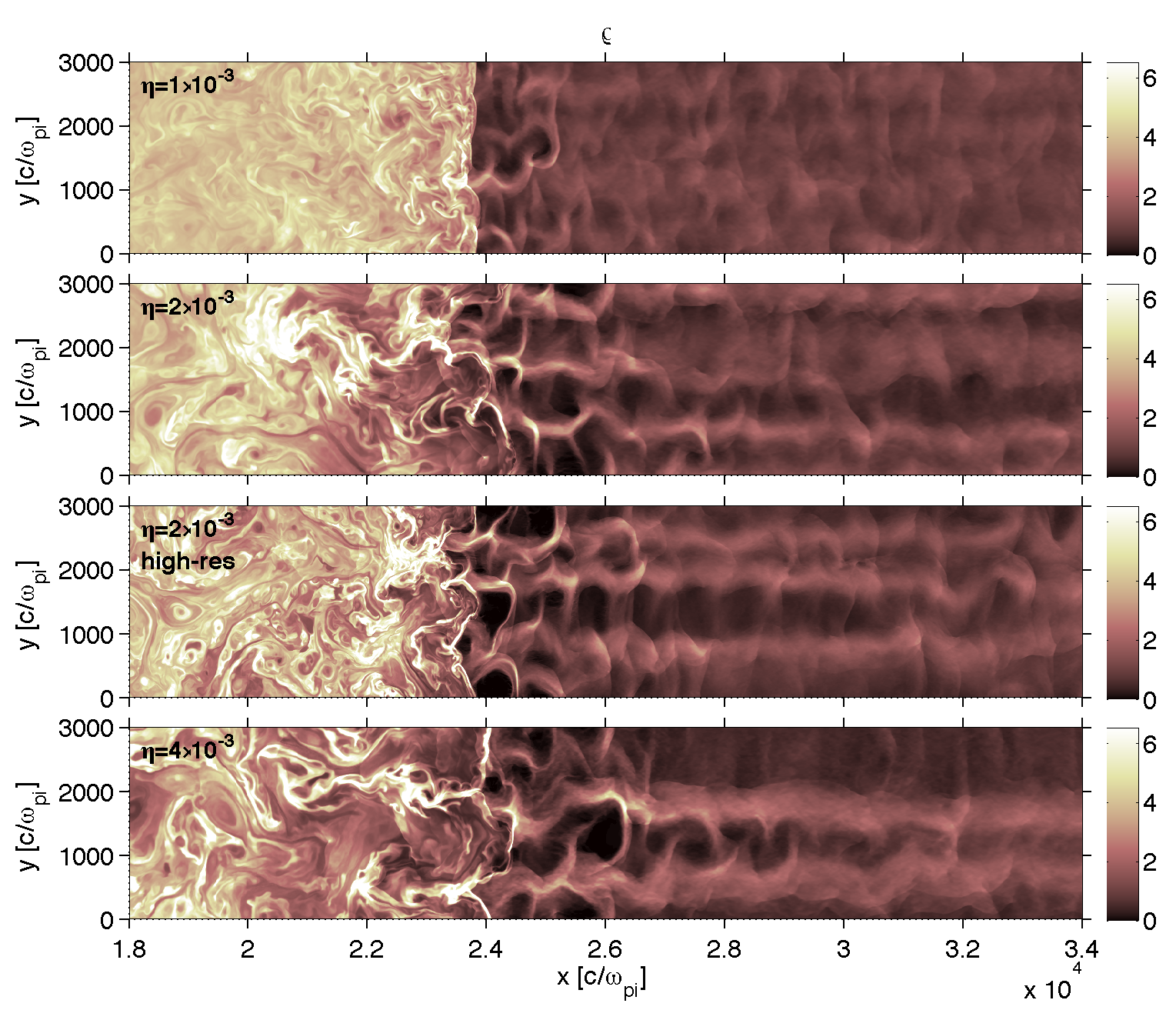}
     \includegraphics[width=89mm]{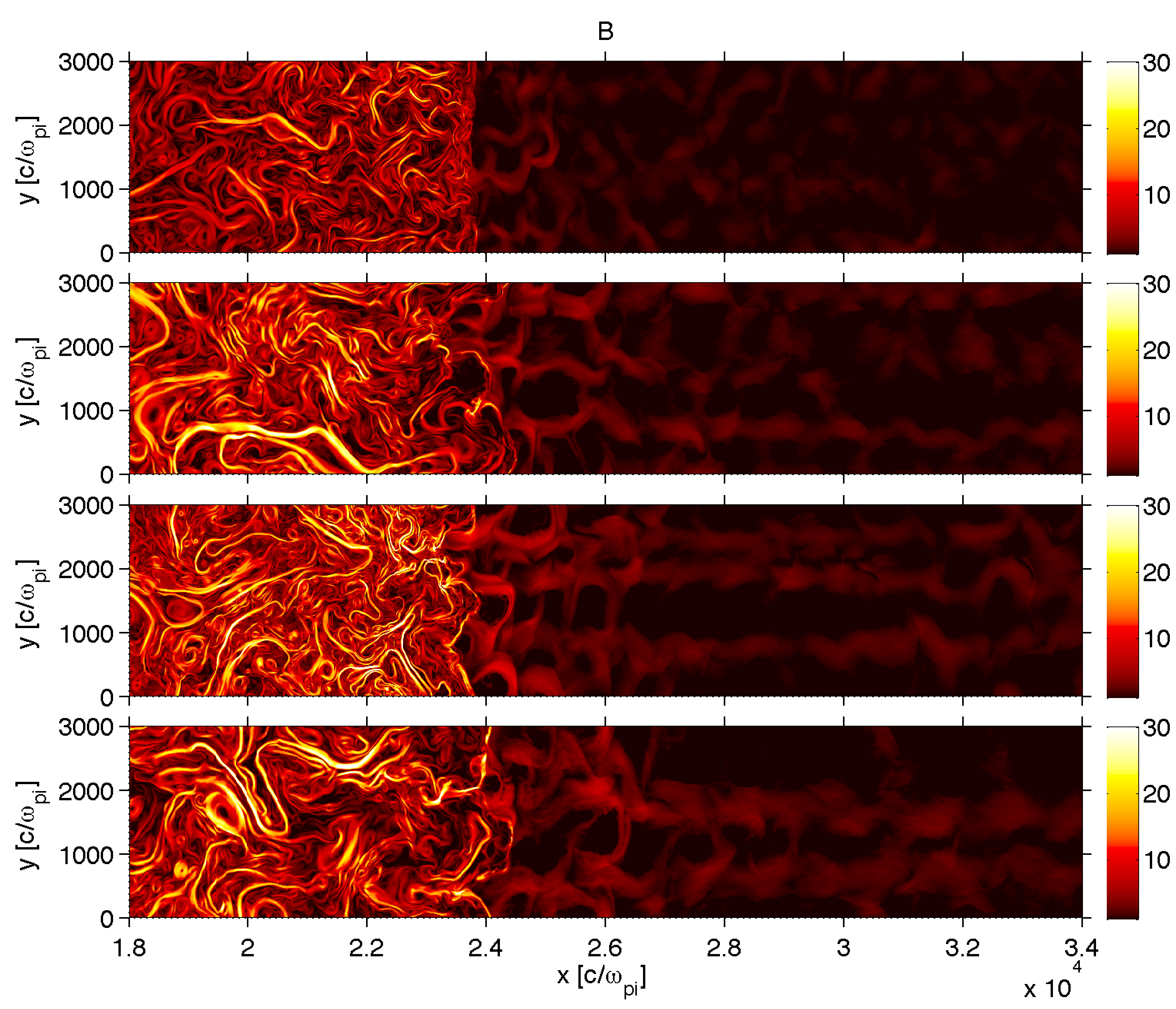}
  \caption{Simulation snapshots of gas density (left) and total magnetic field
  strength (right) at $t=2400\Omega_c^{-1}$ from four of our simulation runs
  R1, R2, R2-hr and R4 with increasing injection efficiency $\eta$
  as labeled in each panel. The third panels from top correspond to the high
  resolution run. The shock is approximately located at $x=24000$.}\label{fig:snapshot}
\end{figure*}

\section[]{Particle Acceleration in Collisionless Shocks: Non-relativistic Regime}
\label{sec:results}

We focus on runs with ${\mathbb C}\gg v_0$ in this section, where all particles
remain non-relativistic at all times. We begin by discussing simulation results from
our fiducial run R2 in Sections \ref{ssec:struct} and \ref{ssec:acc}, followed by
parameter study in Section \ref{ssec:param}.

\subsection[]{Shock Evolution and Structure}\label{ssec:struct}

In our simulations, the Bell instability is first excited in the shock upstream
due to the streaming CRs, and exhibits clear signature of circularly polarized
modes in transverse magnetic and velocity fields, as expected from the linear
eigenvector shown in Appendix \ref{ssec:bellgrow} (Equation
\ref{eq:eigenvect}). Growth into the non-linear stage leads to strong density
variation and magnetic field amplification. In Figure \ref{fig:evolve}, we show
snapshots of gas density at four different times from our fiducial run R2.
Note that only a small fraction ($15\%$) of the simulation box in the vicinity of
the shock front is shown. While the Bell instability is essentially
incompressible in the linear regime, the gas is gradually evacuated to produce
cavities towards the non-linear regime as a result of the filamentation
instability \citep{Bell05,RevilleBell12}. The density structure found here is
qualitatively very similar to that found in hybrid-PIC simulations of
\citet{CaprioliSpitkovsky13}, justifying the validity of our MHD-PIC approach.
We note that over relatively long time evolution, the system develops larger
and larger structures, which can only be accommodated with large transverse
simulation domain size adopted in our simulations (as opposed to typical domain
size of $\lesssim1000c/\omega_{pi}$ in hybrid-PIC simulations).

\begin{figure}
    \centering
    \includegraphics[width=90mm]{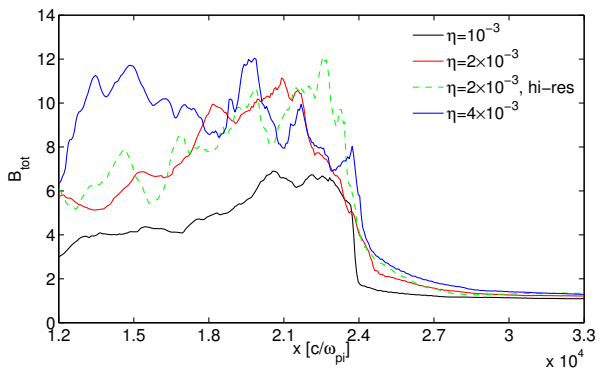}
  \caption{Transversely averaged profiles of magnetic field strength in the vicinity
  of the shock front around time $t=2400\Omega_c^{-1}$ for four of our
  simulation runs R1, R2, R2-hr and R4.}\label{fig:Bprof}
\end{figure}

In Figure \ref{fig:snapshot}, we further show snapshots of the gas density and
magnetic field strength at time $t=2400\Omega_c^{-1}$ from all our simulations
with non-relativistic particles (all runs except R2-REL). We focus on our fiducial
run (second row) in this section. Accompanied by the Bell instability with
cavitation and filamentation, we find that upstream magnetic fields are amplified
by a factor of up to $\sim3-5$ in the strongest regions (corresponding to density
filaments), which is achieved gradually as the flow approaches the shock. Once
processed by the shock, magnetic fields are amplified further due to compression
and vorticity generation from the shock interface (e.g., Richtmeyer-Meshkov
instability), and develop into filamentary structures of strongly enhanced fields a
factor of up to 30 times the background field strength. All these features are very
similar to results from self-consistent hybrid-PIC simulations of \citet{CaprioliSpitkovsky13}.

Figure \ref{fig:Bprof} further shows the profile of mean field strength in the
vicinity of the shock front, where we have smoothed the profiles by averaging from
time $t=2300\Omega_c^{-1}$ to $t=2500\Omega_c^{-1}$ (with proper positional shift
to account for shock motion). Focusing on run R2 here,
we see that prior to the shock front, the mean field strength has increased
by a factor of $\sim2$. The downstream mean field strength is about $6-10$
stronger than the initial field $B_0=1$, which is again consistent with hybrid-PIC
simulations of \citet{CaprioliSpitkovsky14a}. We also note that while the
downstream profile of mean field strength is highly time variable, which is also
reflected in its large spatial variations, the profile shown in the Figure is rather
typical.

\begin{figure}
    \centering
    \includegraphics[width=90mm]{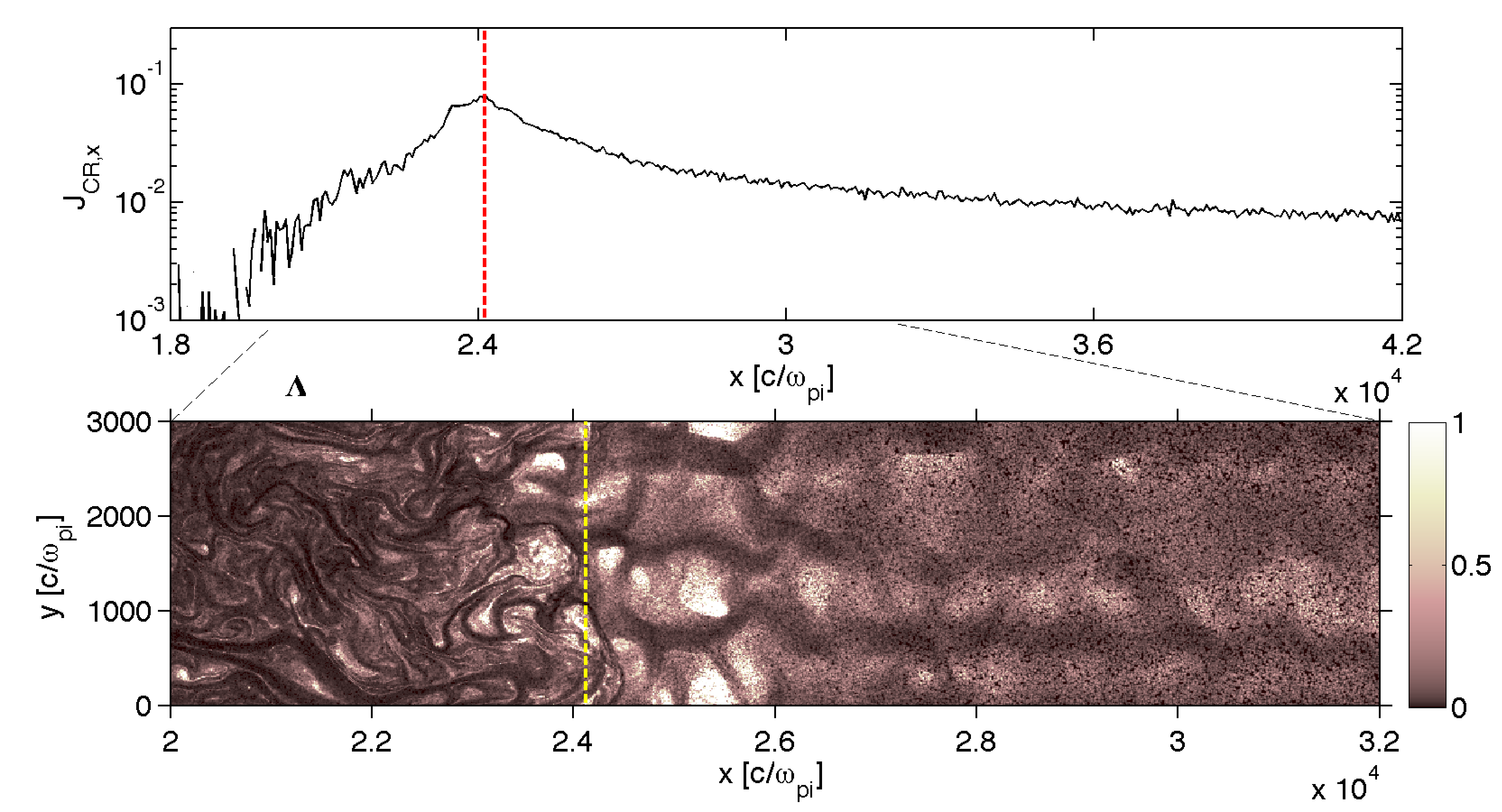}
  \caption{Results from our fiducial simulation R2 at $t=2400\Omega_c^{-1}$.
  Top: transversely averaged $x$-component of the CR current density profile
  along $\hat{x}$ in the co-moving frame. Bottom: snapshot of
  $\Lambda$ defined in Equation (\ref{eq:lambda}) at $t=2400\Omega_c^{-1}$,
  which measures the importance of the CR-Hall term. Vertical dashed
  lines in the two panels indicate the location of the shock front.}\label{fig:CRHsim}
\end{figure}

We next show, in the upper panel of Figure \ref{fig:CRHsim}, the transversely
averaged profile of longitudinal CR current density $j_{{\rm CR}, x}$. It is measured in
the co-moving frame of the gas (based on transversely averaged gas velocity in $\hat{x}$),
and is the source of free energy in the Bell instability.
The measured value of $j_{{\rm CR}, x}$ slowly decreases towards the far upstream
of the shock to the level of $<0.01$ in code units, a factor of $8$ smaller than
the initial injected CR current density $(4/3)M_A\eta\sim0.08$. This is because most
of the particles are scattered back downstream within a short distance (diffusion
length) from the shock front, and the far-upstream CR current is carried by escaping
particles accelerated to much higher energies whose number density is much smaller
(see the top panels of Figure \ref{fig:spectraT} discussed in the next subsection for
more information). These features are all consistent with hybrid-PIC simulations
\citep{CaprioliSpitkovsky14b}. We can infer the most unstable wavelength for
the Bell instability away from the shock front to be
$\lambda_m\sim4\pi B_0/j_{{\rm CR}, x}\sim1.2\times10^3(c/\omega_{pi})$, which is
very well resolved in our simulations. This scale is also consistent with the typical
scale of density and magnetic fluctuations in the shock upstream present in Figure
\ref{fig:snapshot}. Toward the shock front, the value of $j_{{\rm CR}, x}$ increases
to its peak value of near $0.1$, which can be considered as the sum of injected
CR current and the current from high energy CRs. It drops quickly to about zero in
the downstream, since no relative motion between gas and CRs is expected.

In Sections \ref{sec:eqs} and \ref{sec:crcdi}, we have highlighted the potential
importance of the CR-Hall term and its relevance to the Bell instability in
collisionless shocks.
In the bottom panel of Figure \ref{fig:CRHsim}, we further show the map of
$\Lambda$ defined in Equation (\ref{eq:lambda}) at $t=2400\Omega_c^{-1}$,
which measures the importance of the CR-Hall term. The spatial distribution of
$\Lambda$ is very non-uniform. Comparing with the corresponding panels in
Figure \ref{fig:snapshot}, we see that $\Lambda$ is the largest in the density cavities,
and achieves order unity near the shock front. We also see that at up to $\sim10^3$
ion inertial lengths ahead of the shock, $\Lambda$ already has non-negligible
deviations from zero. This result justifies the necessity of including the
CR-Hall term to capture the role of streaming CRs on the evolution of
magnetic fields. On the other hand, since $\Lambda\propto j_{{\rm CR}, x}$
by definition, we have $\Lambda\ll1$ toward the far upstream (not shown in the
Figure), and the CR-Hall effect becomes negligible.

\subsection[]{Particle Acceleration}\label{ssec:acc}

Since we inject particles with sufficiently high energy $E_k=10E_{sh}$
into the shock upstream, most of the injected CR particles directly enter the DSA
process. As long as the scattering process is isotropic in
the co-moving frame, the expected spectrum of accelerated particles does not
depend on the details of the scattering, but only on the compression ratio $r$,
with the momentum spectrum of accelerated particles to be $f(p)\propto p^{-q}$,
where $q=3r/(r-1)\simeq4$ for $r\simeq4$. The energy spectrum is related to
the momentum spectrum via
\begin{equation}
f(E)=4\pi p^2f(p)\frac{dp}{dE}\ .\label{eq:dpdE}
\end{equation}
For non-relativistic particles considered here, the energy spectrum is expected
to have the form $f(E)\propto E^{-3/2}$.

\begin{figure}
    \centering
    \includegraphics[width=85mm]{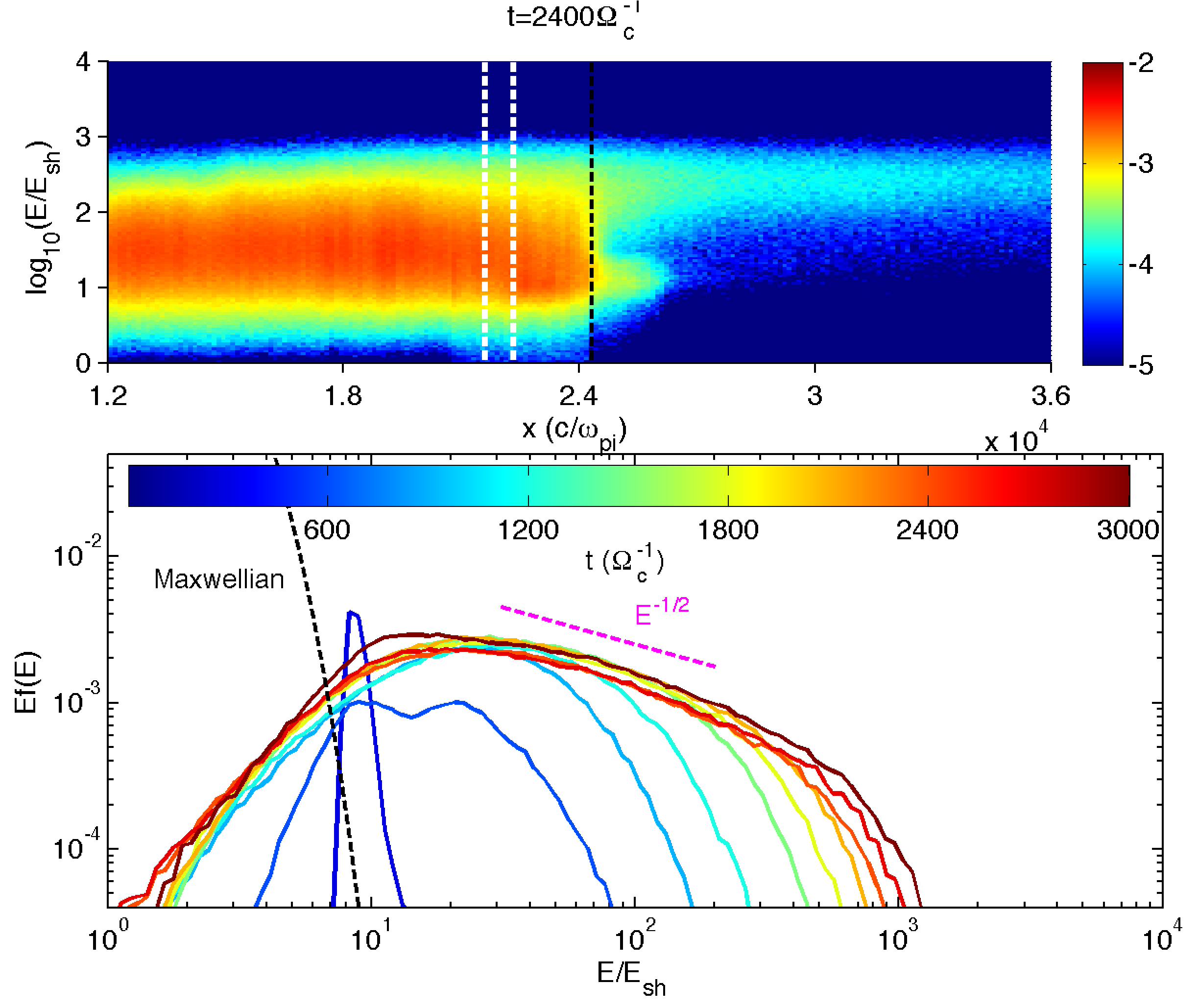}
  \caption{Energy spectrum (in dimensionless form $Ef(E)$) of the CR particles from
  our fiducial run R2. The top panel shows the energy spectrum as a function of
  $x$ in units of $E_{sh}=v_0^2/2$ at time $t=2400\Omega_c^{-1}$.
  The bottom panel shows the time evolution of the downstream particle energy
  spectrum, as marked with different colors indicated by the color table.
  The spectrum is extracted by averaging through a layer at a fixed distance
  behind the shock front. As an example, vertical black dashed lines in the top
  panels mark the location of the shock front, and the vertical white dashed
  lines mark the layer where the downstream spectrum is extracted. The black
  dashed line shows the thermal energy spectrum with temperature
  $T=0.85T_1$.}\label{fig:spectraT}
\end{figure}

In the bottom panel of Figure \ref{fig:spectraT}, we show the time evolution of
the downstream particle energy spectrum for our fiducial run R2. At early
times before turbulence is fully developed ($\lesssim300\Omega_{c}^{-1}$), the
particle spectrum reflects the initial injected energy distribution, which peaks at
$E=10E_{sh}$. The spectrum then significantly broadens, extending
substantially to the high energy side with a very small fraction scattered to low
energy side. A high-energy power-law tail is gradually built up, with spectral
slope consistent with ${-3/2}$. The high-energy tail extends to higher and higher
energies with time, with a roughly exponential energy cutoff. We note that the
normalization of the downstream spectrum appears to vary slightly in time and
space, which reflects the stochasticity of the shock. For example, by performing
additional simulations with different random seeds, we quote a relative uncertainty
of $\sim20\%$ in the normalization.

We also show in black dashed line the expected energy spectrum for a
thermal particle distribution (Maxwellian) in the downstream
\begin{equation}
Ef(E)=4\times(2/\sqrt{\pi})(E/E_{\rm th})^{3/2}e^{-E/E_{\rm th}}\ ,
\end{equation}
where we choose $E_{\rm th}=0.85T_1$ (note that  $T_1=v_0^2/3$ is the downstream
temperature from the shock jump condition). The deduced $15\%$ serves as a
proxy for the fraction of kinetic energy converted to accelerate particles,
and the pre-factor of $4$ is to account for shock compression. Clearly, the
Maxwellian distribution is distinct from our CR energy distribution as they can
not be smoothly joined together. In reality, as shown in self-consistent hybrid-PIC
simulations, the energy spectrum is smooth across the entire energy range thanks
to a population of supra-thermal particles whose energy lies between a few $E_{sh}$
to about $10E_{sh}$ \citep{CaprioliSpitkovsky13}. The presence of the supra-thermal
particles is related to the particle injection process that bridges the gap between
thermal and non-thermal particle populations \citep{Caprioli_etal15}. While this piece of
physics is currently missing in our simple injection prescription, our current
prescription may serve as a first approximation to mimic the injection process, with
injected $E=10E_{sh}$ particles presumably fed by the supra thermal particle population.
We plan to develop and employ more realistic injection prescriptions in the near future.

In the upper panel of Figure \ref{fig:spectraT}, we show the spatial distribution of the
particle energy spectrum at relatively late stage of the shock with
$t=2400\Omega_c^{-1}$. We see that high-energy CRs with energies $\sim100E_{sh}$
or higher penetrate into the shock upstream and provide the source of CR current to
drive the Bell instability. 
The lower protrusion at $E_k\sim10E_{sh}$ into the shock upstream is mainly due to
the artificial injection procedure adopted here, and we see that the initially injected CR
particles are mostly confined in regions close to the shock front. 
We measure the particle energy spectrum at a fixed distance of about
2400$c/\omega_{pi}$ behind the shock, as indicated in the Figure.
As discussed in \citet{CaprioliSpitkovsky14a}, this allows the shape of the particle energy
spectrum to be fully settled.

We have also examined the evolution of the maximum particle energy.
While we can always identify the particle with the maximum energy, the time evolution
of this energy can be rather noisy due to random scattering and due to rapid changes
when such particle leaves the domain. Here, we instead consider the following
definition based on the full energy spectrum
\begin{equation}\label{eq:emax}
\bar{E}_{\rm max}=\frac{\int E^{n+1}f(E)dE}{\int E^{n}f(E)dE}\ ,
\end{equation}
where $n$ is an integer number of our choice. Note that if the energy
distribution function takes the form $f(E)\propto E^{-m}\exp(-E/E_{\rm cut})$,
then this integral yields $\bar{E}_{\rm max}\approx(n+1-m)E_{\rm cut}$.
Empirically, we choose $n=6$, which corresponds to $\bar{E}_{\rm max}$
being $\sim5-5.5E_{\rm cut}$ if the cutoff is exponential. We also find that the
number obtained this way is approximately half the absolutely maximum particle
energy.

\begin{figure}
    \centering
    \includegraphics[width=90mm]{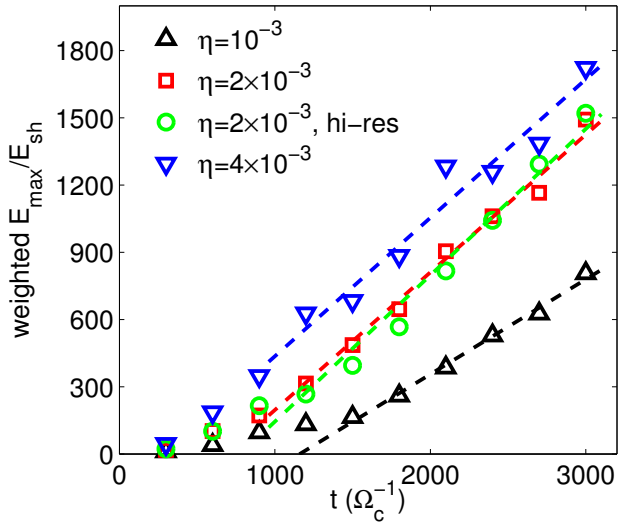}
  \caption{Time evolution of the weighted maximum particle energy
  $\bar{E}_{\rm max}$ from four of our simulation runs R1, R2, R2-hr and R4 with
  increasing CR injection efficiency $\eta$. Also shown are linear fits to the data
  with $t>960\Omega_{c0}^{-1}$.}\label{fig:EmaxT}
\end{figure}

The time evolution of $\bar{E}_{\rm max}$ is directly related to the rate
of particle diffusion across the shock, and hence closely probes the properties
of the MHD turbulence self-generated by streaming CRs. This is shown in
Figure \ref{fig:EmaxT}. Excluding the time before $t=960\Omega_{c0}^{-1}$
(which is expected to be affected by initial conditions), we see that
$\bar{E}_{\rm max}$ increases approximately linearly with time. This is
consistent with efficient turbulent diffusion/scattering of particles with diffusion
coefficient $D_p\propto v_pR_L$, where $v_p$ and $R_L$ are particle velocity
and gyro-radius \citep{CaprioliSpitkovsky14c}. 
Being a proof-of-concept study with simplified injection prescription,
performing more quantitative analysis on the properties of magnetic turbulence
and particle diffusion is beyond the scope of this work, but we expect the detailed
analysis performed in \citet{CaprioliSpitkovsky14b,CaprioliSpitkovsky14c} to
hold in general.

Based on the energy spectrum, we can measure the efficiency of particle
acceleration $\xi$ by calculating the fractional energy residing in the CR
particles compared with the shock kinetic energy ($mv_{sh}^2/2$). In the
simulations, we find that this fraction first increases with time, and then
achieves an approximately constant level but still varies with time at
some level. By averaging the measured efficiency over the last
$500\Omega_c^{-1}$ of our simulations, we find $\xi\approx13\%$ from our
fiducial run R2 (with relative uncertainty of $\sim20\%$, as discussed earlier),
which is comparable to the efficiency obtained in hybrid simulations
\citep{CaprioliSpitkovsky14a}.


\subsection[]{Parameter Dependence}\label{ssec:param}

The outcome of the simulations mainly depends on the prescribed CR injection
fraction $\eta$, an artificial parameter of the simulations. Snapshots of
the shock structure for all simulations with non-relativistic particles at
$t=2400\Omega_{c0}^{-1}$ are shown in Figure \ref{fig:snapshot}.
For relatively small injection fraction $\eta=10^{-3}$ (run R1), the CR current
streaming through the upstream, and hence the growth rate of the CR-driven
instabilities, are relatively small. Correspondingly, it takes longer for cavitation
and filamentation to develop, with smaller density contrast before reaching the
shock front. While turbulence is generated in the upstream, the shock profile
remains sharp at all times in our simulation.

When increasing $\eta$, cavitation becomes stronger, leading to higher density
contrast and stronger magnetic fluctuations in the shock upstream. As a result,
the shock front is more disturbed, leading to thicker shock transition layer (which
is more evident in the temperature profile not shown in the plots). For run R4 with
$\eta=4\times10^{-3}$, violent turbulence already dominates the shock near
upstream, and the shock front substantially smoothed but is still identifiable. We
have also performed simulations with $\eta=5\times10^{-3}$, and find that the
shock front is so much disrupted that it becomes difficult to identify the location of
the shock. In realistic shocks, such strong modification of shock structure is expected
to lead to a reduction of the injection efficiency, and we conclude that imposing a
constant $\eta\gtrsim4\times10^{-3}$ leads to an unphysical over-injection.

Different levels of magnetic field amplification in the shock upstream can be clearly
identified in Figure \ref{fig:Bprof}, where the mean field strength near the shock front
on the upstream side increases with increasing $\eta$, as a result of stronger
upstream CR current. Correspondingly, the downstream field strength also
increases with $\eta$ within the duration of our simulations.

\begin{figure}
    \centering
    \includegraphics[width=90mm]{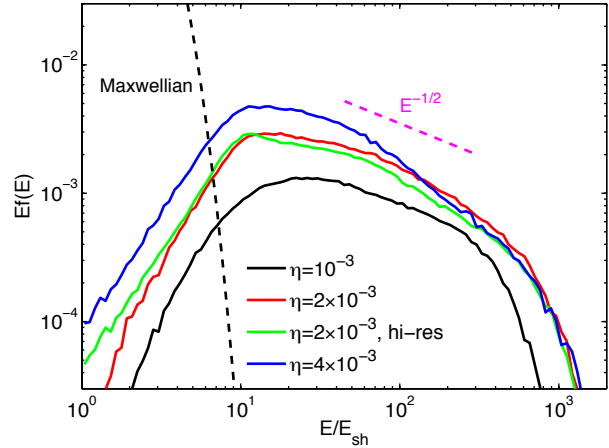}
  \caption{CR particle energy spectrum at $t=3000\Omega_c^{-1}$ in the shock
  downstream from our four simulation runs R1, R2, R2-hr and R4.
  The black dashed line shows the thermal energy spectrum with temperature
  $T=0.85T_1$.}\label{fig:spectraP}
\end{figure}

Particle acceleration is very efficient in all simulations. Figure
\ref{fig:spectraP} shows the cross-comparison of  downstream particle energy
spectrum measured at $t=3000\Omega_c^{-1}$, measured from our four simulations
with increasing $\eta$. For $\eta\lesssim2\times10^{-3}$, the particle energy
spectra follow the $f(E)\propto E^{-3/2}$ profile before the cutoff, while with
relatively large $\eta=4\times10^{-3}$, the spectra slope is less consistent,
especially at higher energies. Clearly, the total energy contained in the CR
particles increases with the injection parameter $\eta$, and we quote
$6\%$ and $20\%$ as the acceleration efficiency $\xi$ from runs with
$\eta=10^{-3}$ and $4\times10^{-3}$, respectively. The former is smaller than
obtained from hybrid-PIC simulations, indicating under-injection, while the latter is
higher, indicating over-injection. The deviation from the expected spectral
shape in run R4 may be the result of over-injection.

In all cases, the maximum particle energy increases approximately linearly with
time, as seen from Figure \ref{fig:EmaxT}. The rate of increase is shallower with
smaller $\eta$, suggesting less efficient particle diffusion. This is in line with the
fact that there is less magnetic field amplification/fluctuation with smaller $\eta$,
which leads to larger diffusion coefficient. Increasing $\eta$ to $4\times10^{-3}$,
the rate at which $\bar{E}_{\rm max}$ grows follows more closely with our fiducial
run, and also shows larger fluctuations. This also reflects the saturation of
magnetic field amplification and particle acceleration efficiency, again
suggestive of over-injection.

Overall, our parameter study suggests that the sharpness/smoothness of the shock
depends on the prescribed particle injection efficiency $\eta$, which also determines
the efficiency of particle acceleration $\xi$. Our fiducial choice of
$\eta=2\times10^{-3}$ appears to yield results most consistent with hybrid-PIC
simulations, at least within the duration of our runs. Over-injection of
particles would over-smooth the shock. Since this is not observed in self-consistent
simulations, particle injection must be suppressed when the shock transition layer
is over-smoothed, so that in reality the shock structure remains reasonably sharp,
and the particle acceleration efficiency should be kept to be within a certain level
$\lesssim20\%$. In this sense, we conclude that particle injection must be a
self-limiting process.

\subsection[]{Convergence}\label{ssec:res}

Our high-resolution run R2-hr is in most aspects very similar to our fiducial run
R2. Comparing their shock structures in Figure \ref{fig:snapshot}, we see that
run R2-hr develops cavities at similar locations relative to the shock front as the
run R2. The density contrast in these cavities and the level of magnetic
field amplification are also similar between the two runs. While the shock transition
region in the high resolution run appears to be thinner than in the low resolution
run {\it at this particular snapshot}, this is mostly due to the very dynamic nature of
the shock with highly chaotic turbulence. Over longer timescales, the two runs
show no significant difference in shock appearance.
More definitive evidence of convergence is revealed in Figures \ref{fig:EmaxT}
and \ref{fig:spectraP}, where we see that the particle energy spectrum (hence
acceleration efficiency), together with the evolution of maximum particle energy,
agree quantitatively between the low and high resolution runs. This in turn implies
that the particle diffusion coefficients are quantitatively similar between the runs.

\begin{figure}
    \centering
    \includegraphics[width=90mm]{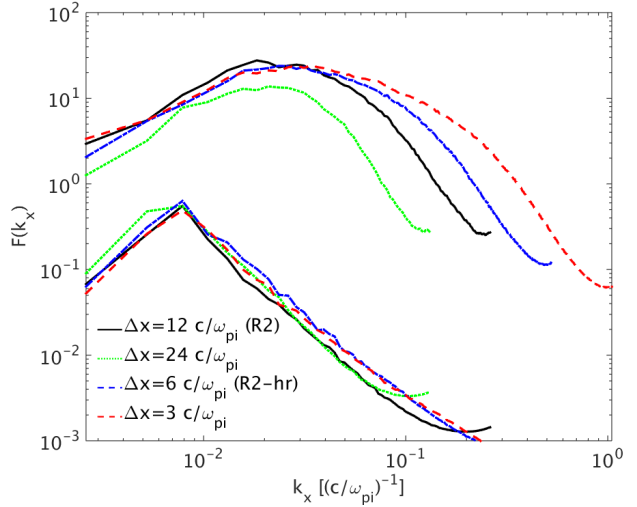}
  \caption{Dimensionless spectral energy distribution $\mathcal{F}(k_x)$ of
  transverse magnetic fluctuations from our resolution study of shock simulations,
  averaged over a time period between 2160 and 2400 $\Omega_{c}^{-1}$.
  Upper lines correspond to downstream spectrum, computed over a distance of
  2400 $c/\omega_{pi}$ in $x$ starting from 2000 $c/\omega_{pi}$
  behind the shock. Lower lines correspond to upstream spectrum, computed over
  a distance of 2400 $c/\omega_{pi}$ in $x$ starting from 4000 $c/\omega_{pi}$
  ahead of the shock. Different colors correspond to different resolutions, as
  indicated by the legend, where black solid lines correspond to our fiducial
  resolution run R2. The rightmost point of each line corresponds to
  grid scale of the respective runs.}\label{fig:pwspec}
\end{figure}

We have also computed the spectral energy distribution ${\mathcal F}(k_x)$
of transverse magnetic fluctuations following the approach of
\citet{CaprioliSpitkovsky14b} (see their Equation (3)), and show the results in
Figure \ref{fig:pwspec}. Here, ${\mathcal F}(k_x)$ is dimensionless and measures
the energy density of magnetic fluctuations per logarithmic wavenumber,
normalized to background field energy density.
To better demonstrate convergence, we performed two additional runs with half
the resolution of R2 and twice the resolution of R2-hr, and hence in total we have
four runs with resolution spanning from 24 to 3 $c/\omega_{pi}$ per cell. We
compute ${\mathcal F}(k_x)$ at fixed distances at the upstream and downstream
sides of the shock for each run (see the caption of Figure \ref{fig:pwspec} for details),
averaged over time interval between 2160 and 2400 $\Omega_{c}^{-1}$.

We see that in the upstream, the Bell instability is well resolved at all resolutions
(modulo uncertainties in the overall normalization as discussed previously). In all
cases, the spectrum peaks at $k_x\sim0.008(c/\omega_{pi})^{-1}$, corresponding
to the fastest growing Bell mode ahead of the shock\footnote{This corresponds
to longer wavelength than the naive estimate in Section \ref{sec:shock}, since most
injected particle are scattered back shortly without reaching large distance into the
upstream.}. In the downstream, the lowest resolution simulation have somewhat
less power and may be under-resolved (but can also be due to larger noise and
variability since there are fewer resolution elements compared to higher resolution
runs). For our fiducial run R2, the power spectrum matches well with higher resolution
runs at small $k_x$.\footnote{While the spectra do not line up exactly at lowest
$k_x$, this is most likely due to small number statistics of long-wavelength modes.}
Increasing resolution, the spectrum extends to higher $k_x$. We note that for
typical grid-based MHD code, $\sim16$ cells are required to properly resolve a
single wave mode without strong numerical dissipation (e.g., see Figures 7
and 12 of \citealp{Stone_etal08} and discussions therein). This is exactly
the scale where the power spectrum of our fiducial run R2 starts to deviate from
higher-resolution runs at $k_x\sim0.03(c/\omega_p)^{-1}$. We also see from the
highest resolution run that interestingly, the downstream spectrum peaks at a scale
about four times smaller than the scale of the upstream spectral peak. This may be
simply understood as a result of shock compression.

Overall, since CR acceleration relies on the turbulent power at larger scales,
which are well resolved in all runs,
this leads to consistent calculation of the particle spectrum and of the maximum energy
achieved by accelerated particles (see Figures \ref{fig:EmaxT} and \ref{fig:spectraP}).
Our convergence study thus gives us confidence that our fiducial resolution of
$12(c/\omega_{pi})$ per cell is sufficient to capture the essential physics.

\section[]{Particle Acceleration in Collisionless Shocks: Transition to Relativistic Regime}\label{sec:trans}

In this section, we discuss our last run R2-REL, where we use a reduced particle
speed of light ${\mathbb C}=10\sqrt{2}v_0$ to study particle acceleration with transition
from non-relativistic to relativistic regime. Note that the initial particle energy of
$E_k=10E_{sh}$ corresponds to typical particle velocity of
$v\sim3.2v_0\sim0.22{\mathbb C}$. The corresponding $\gamma\approx1.025\approx1$,
hence particles can still be considered as non-relativistic. Transition to the relativistic
regime occurs approximately when $E_{k,{\rm trans}}\sim{\mathbb C}^2/2=200E_{sh}$,
where $\gamma\approx1.5$. This way, we expect the particle energy spectrum to span
nearly one decade before becoming relativistic. We run the simulations sufficiently long to
achieve another decade of energy span in the relativistic regime\footnote{The time we
terminate the simulation corresponds to the maximum time when the flow that reaches the
shock front is unaffected by the boundary condition at the rightmost end of the box. Since
it takes time for the accelerated particles to reach the right boundary, this time is larger than
$3L_x/4v_0=9720\Omega_c^{-1}$.}.

\begin{figure*}
    \centering
    \includegraphics[width=180mm]{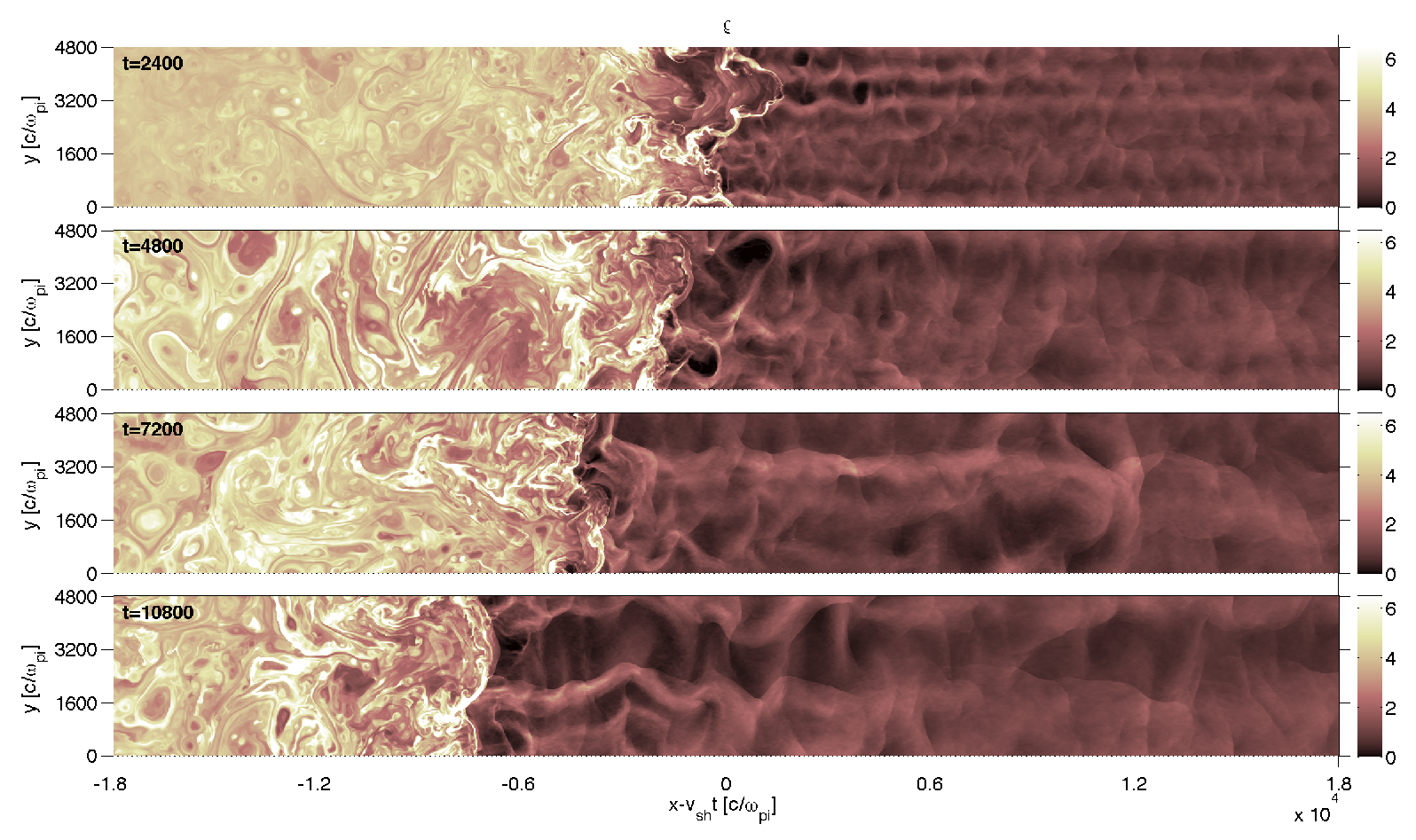}
  \caption{Snapshots of gas density at four different times (as indicated) from
  our run R2-REL with reduced speed of light ${\mathbb C}=10\sqrt{2}v_0$ at
  four different times as labeled in each panel (unit is $\Omega_c^{-1}$). Only
  a small fraction of the simulation box is shown which is adjusted to cover the
  vicinity of the expected shock location at $x=v_{sh}t=v_0t/3$.}\label{fig:evolve_re}
\end{figure*}

\subsection[]{Shock Structure and Evolution}

In Figure \ref{fig:evolve_re}, we show the time evolution of the gas density in
the vicinity of the shock front. While the overall sequence shares similarities with
our fiducial run R2 shown in Figure \ref{fig:evolve}, there are several points worth
further discussion.

First, we clearly see that the waves and filamentation/cavity
structure in the shock upstream progress to larger sizes over time. This is the main
reason that we have enlarged our box size to $4800(c/\omega_{pi})$ to accommodate the
growing structures. Near the end of the run (bottom panel), the size of the largest
structure is already a significant fraction of the transverse box size. Our additional
test simulations with smaller transverse box size tend to saturate  into a state with
a single dominant transverse mode in the shock upstream, which may affect the
scattering of energetic CR particles.


Second, we notice that the shock upstream in this run appears to be less perturbed and has a 
sharper shock transition layer compared to our fiducial run R2 (e.g., at time
$t=2400-3000\Omega_c^{-1}$). This is mainly due to the use of a smaller speed of light than
in run R2. Note that the turbulence in the shock upstream is mainly excited by the escaping CRs,
whose contribution is mainly due to the net CR current density. However, with a reduced
speed of light, the CR velocity saturates at a smaller value of ${\mathbb C}$: higher-energy CR particles
become relativistic and do not contribute as much current as they do as non-relativistic
particles. This effect becomes more pronounced with time, since the particles that
penetrate into the shock upstream have higher energies.

Third, we notice that at later times the shock velocity slightly deviates from the standard jump condition of $v_{sh}=v_0/3$, dropping by about $10\%$ from time $t=4800\Omega_c^{-1}$ to $10800\Omega_c^{-1}$. 
As a result, the shock front lags from the expected position (Figure \ref{fig:evolve_re}), and the shock compression ratio density is $r\approx 4.3$, slightly larger than expected from standard jump condition (Equation \ref{eq:rh}). 
This is a direct consequence of channeling of a sizable fraction of the upstream ram pressure into accelerated particles that do not feel the standard entropy jump at the shock, as discussed in \S 6.2 of \citet{CaprioliSpitkovsky14a}.
When relativistic CRs contain a sizable fraction of the energy density, a further increase of the compression ratio is expected because of the contribution of relativistic CR fluid (whose adiabatic index is 4/3), but this effect is still negligible in the simulations presented here. 
We also find the development of a CR-induced precursor, in which the pre-shock fluid is slowed down and heated up  because of the pressure in diffusing CRs and because of turbulent heating \citep[see Section 6.1 of][]{CaprioliSpitkovsky14a}.
Note that both the shock precursor and the increase of shock compression ratio are also observed in our previous non-relativistic runs.


\begin{figure}
    \centering
    \includegraphics[width=85mm]{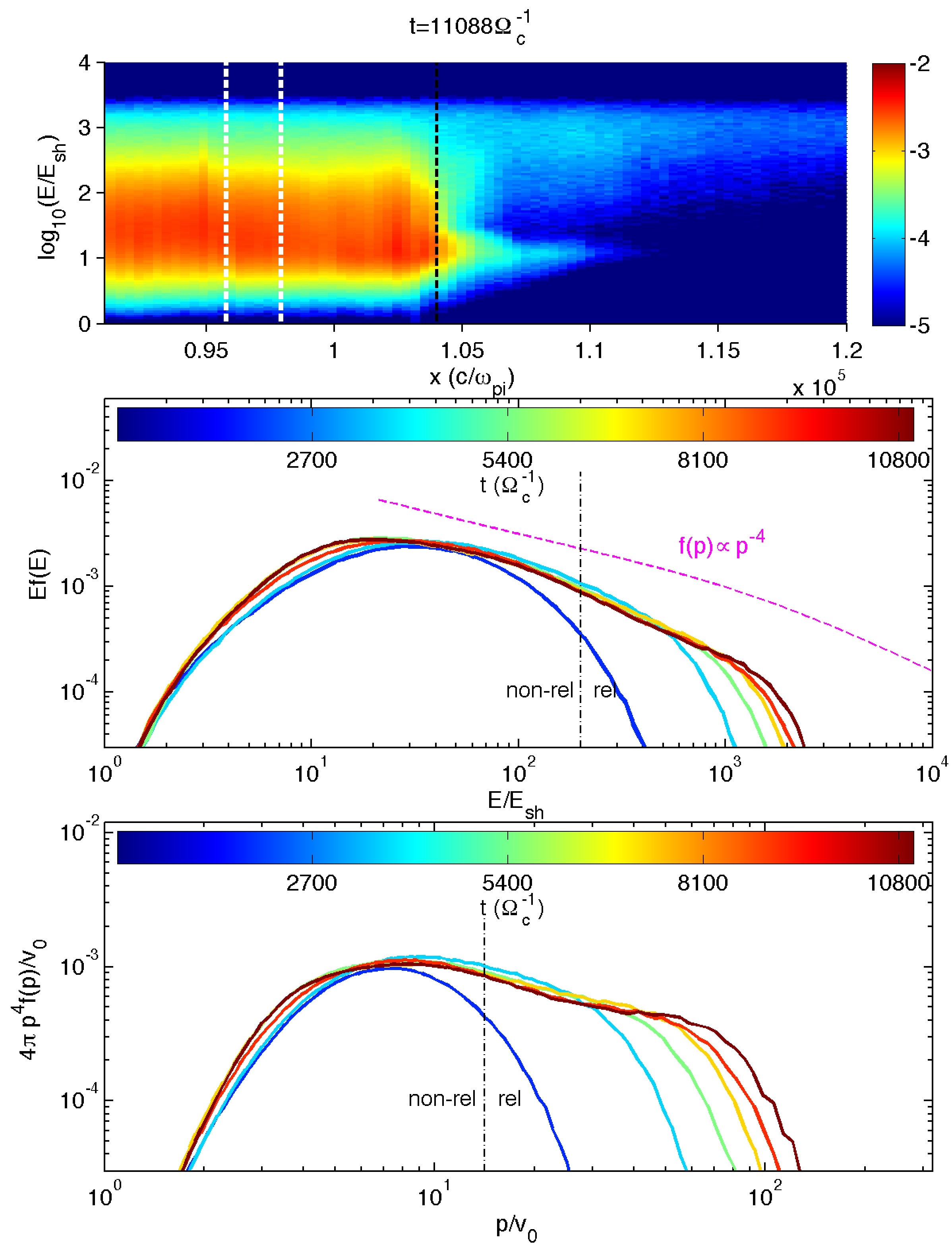}
  \caption{Energy and momentum spectrum of the injected CR particles from
  our run R2-REL with reduced speed of light ${\mathbb C}=10\sqrt{2}v_0$. The top
  panel shows the energy spectrum as a function of $x$ at time $t=11088\Omega_c^{-1}$
  (about the end of the run) in units of $E_{sh}=v_{sh}^2/2$. The middle (bottom) panel
  shows the time evolution of the downstream particle energy (momentum) spectrum as
  marked with different colors indicated by the color table. The energy and momentum
  spectra are shown in dimensionless form $Ef(E)$ and $4\pi p^4f(p)/v_0$), respectively.
  The spectrum is extracted by averaging through a layer at a fixed distance behind
  the shock front. As an example, vertical black dashed lines in the top panels mark
  the location of the shock front, and the vertical white dashed lines mark the layer
  where the downstream spectrum is extracted. The dash-dotted lines in the middle
  and bottom panels mark the transition from non-relativistic to relativistic regimes.
  In the middle panel, the expected energy spectrum from a $f(p)\propto p^{-4}$ momentum
  spectrum is also shown.}
  \label{fig:spectraT_re}
\end{figure}

\begin{figure}
    \centering
    \includegraphics[width=90mm]{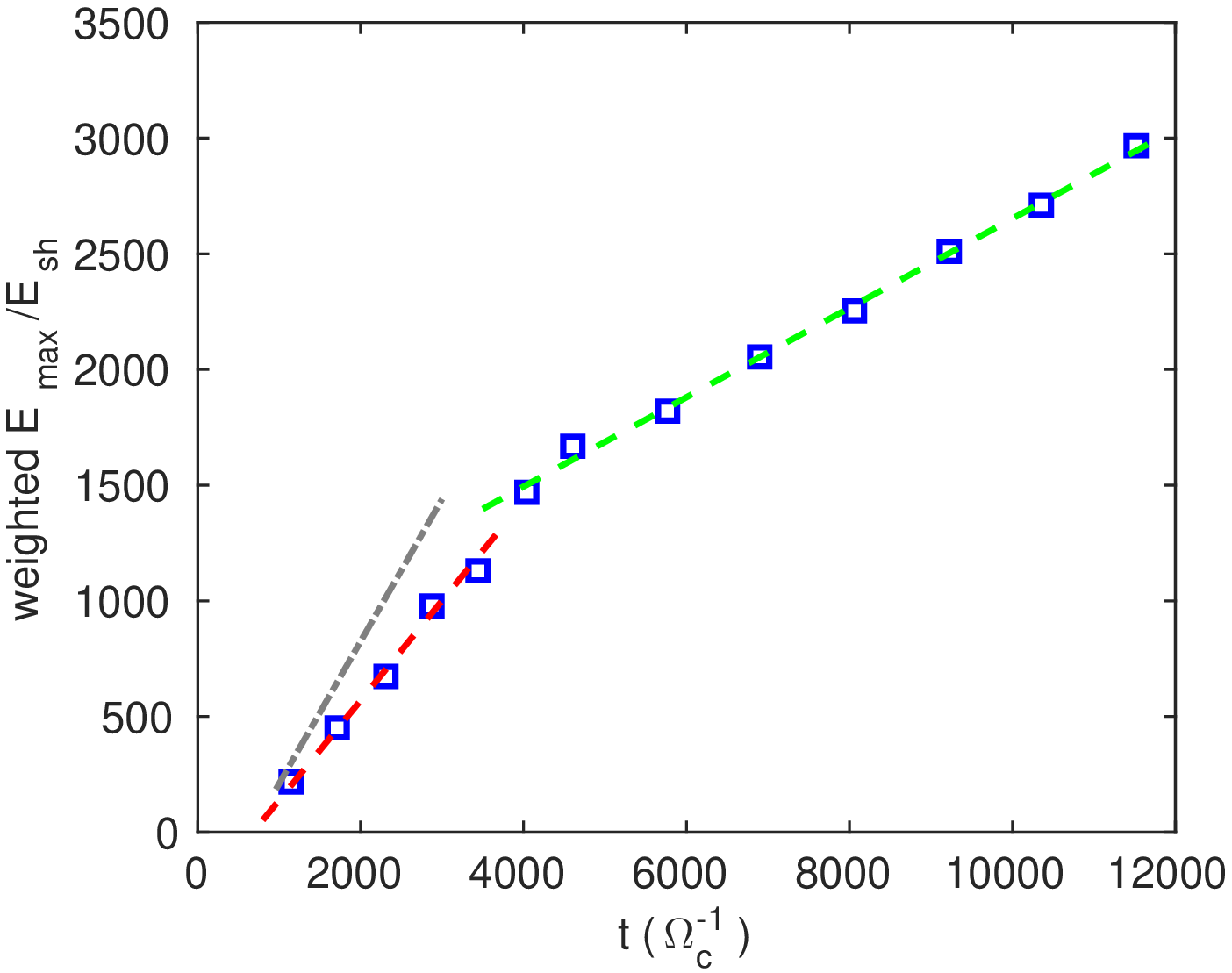}
  \caption{Time evolution of the weighted maximum particle energy $\bar{E}_{\rm max}$
  from our relativistic run R2-REL. Also shown in dashed lines are linear fits to data at
  $t<3000\Omega_c^{-1}$ (red) and $t>4000\Omega_c^{-1}$ (green). The black
  dash-dotted line is a reproduction of the fitting result to run R2 in Figure \ref{fig:EmaxT}:
  acceleration is slower in the relativistic case.}\label{fig:EmaxT_re}
\end{figure}

\subsection[]{Particle Acceleration}

The CR spectrum at the end of the simulation together with its time evolution
are shown in Figure \ref{fig:spectraT_re}.
Since we expect the particle momentum spectrum $f(p)\propto p^{-4}$ to be a
universal consequence of Fermi acceleration regardless of being in the
non-relativistic or relativistic regimes, we show the time evolution of
the particle momentum spectrum in the bottom panel of the Figure.
Given our definition of particle kinetic energy $E_k$ from Equation (\ref{eq:parEk}), and
using Equation (\ref{eq:dpdE}), particle momentum spectrum is related to energy
spectrum by
\begin{equation}
4\pi p^3f(p)=Ef(E)\frac{{\mathbb C}^2p^2}{E(E+{\mathbb C}^2)}\ .
\end{equation}
We see that the momentum spectrum, plotted in dimensionless form
$4\pi p^4f(p)/v_0$, exhibits approximately flat shape both in the non-relativistic
regime with $p/v_0<14.2$, and in the relativistic regime toward higher momenta
$p/v_0\gtrsim20$, consistent with standard theory of Fermi
acceleration.\footnote{While we also expect the spectral slope to slightly
deviate from standard value due to the change in shock compression ratio,
the deviation is at the level of $0.1$, which is not distinguishable due to
noise in the particle energy and momentum spectra.} The
relativistic part of the spectrum has slightly smaller normalization than the
non-relativistic part, while they connect smoothly through the transition.
We also show the energy spectrum in the middle panel of the Figure. Note that
for a momentum spectrum of $f(p)\propto p^{-4}$, the energy spectrum should
not clearly transition to the $f(E)\propto E^{-2}$ scaling until Lorentz factor
$\gamma\gtrsim10$ ($\sim2000E_{sh}$). Since our CR energy spectrum already
drops off at $E\gtrsim1500E_{sh}$, we do not yet cover the fully relativistic portion
of the spectrum.


In the top panel of Figure \ref{fig:spectraT_re}, we see that CRs escaping into the
far upstream near the end of the run almost completely consist of relativistic particles
whose energy is well above $200E_{sh}$. We also find that more contributions
come from non-relativistic particles toward earlier times. These observations are
consistent with the reduction of $j_{{\rm CR},x}$ with time discussed earlier and
is mainly responsible for the weaker turbulence in the shock upstream.

From the simulation, we find that the acceleration efficiency $\xi$, slowly
increases at early time, and saturates to about $20\%$ in the later half of the
run. The relatively large acceleration efficiency is in line with the reduction of 
shock velocity discussed earlier. We also note that the value of $\xi$ in this
run is about $15\%$ at $t\sim3000\Omega_c^{-1}$, comparable to $13\%$
from our fiducial run R2.

In Figure \ref{fig:EmaxT_re}, we show the evolution of weighted maximum
particle energy. We first note that acceleration is slower than our
fiducial non-relativistic run R2 (indicated in black dash-dotted line).
This is most likely due to the reduction in scattering efficiency as
particle transitions into relativistic regime (e.g., smaller gyro-frequency,
and gyro-radius increases with energy more quickly), resulting in slower
acceleration. Since the threshold energy of $E=200E_{sh}$ is achieved at very
early stage of the simulation with $t<1000\Omega_c^{-1}$, the bulk of the growth
in $E_{\rm max}$ occurs in the relativistic regime (note that our definition of
$E_{\rm max}$ is a factor of several larger than the cutoff energy). In addition,
we find that $E_{\rm max}$ increases with time faster at early times
($t\lesssim3000\Omega_c^{-1}$), but becomes slower later. We
therefore provide two linear fits for $t<3000\Omega_{c}^{-1}$ and
$t>4000\Omega_{c}^{-1}$ respectively, where the slopes differ by a factor of $\sim2$.
The difference most likely related to the reduction of CR current as escaping
particles gradually become dominated by relativistic particles, which
in turn affects the strength of the turbulence.
While more detailed analysis of particle diffusion is beyond the scope of
this work, we have performed further experiments with larger and smaller
particle speed of light, and confirm the effects discussed above.

\section[]{Summary and Discussion}\label{sec:conclude}

In this paper, we have rigorously derived an MHD formulation which describes the
interaction between collisionless CRs and a thermal plamsa. Backreaction from
CRs to the thermal plasma is incorporated in the form of momentum, energy and
electromagnetic feedback, as Equations (\ref{eq:gasmotion}), (\ref{eq:engeq})
and (\ref{eq:emfcr2}). While momentum and energy feedback have been well
understood as a result of momentum and energy conservation in the composite system
of CRs and gas, we point out the previously neglected electromagnetic feedback,
which we term as the CR-induced Hall (CR-Hall) term. It becomes important when the
relative drift velocity between electrons and ions induced by CR streaming
approaches the Alfv\'en velocity, or their ratio $\Lambda$ defined in
(\ref{eq:lambda}) approaches order unity. Our formulation is applicable (i) on
scales much larger than the ion inertial length $c/\omega_{pi}$, which is the scale that
MHD is considered applicable to describe thermal plasma, and (ii) when CRs
constitute a negligible fraction of the total gas mass (which is almost always the case in
real systems). While our formulation remains valid when $\Lambda\gtrsim1$ (i.e.,
extremely strong CR current), additional plasma effects may take place that are
not captured in the formulation.

We have implemented this formulation in the Athena MHD code, where CRs are treated as 
kinetic particles using the PIC technique. An artificial speed of light ${\mathbb C}$
can be specified for CR particles so that they can be either non-relativistic or
fully relativistic, as long as ${\mathbb C}$ is much larger than any MHD velocity.
All ingredients of our MHD-PIC code are well tested with carefully designed problems
and have achieved excellent performance. Note that for conventional full PIC or
hybrid-PIC codes, the requirements to accommodate the large Larmor radius and
diffusion length of the CR particles and to resolve microphysical plasma scales (on the
order of the ion or electron inertial length) become
increasingly incompatible with higher CR energy. By circumventing the tiny plasma scales,
our MHD-PIC approach enables us to study the physics of  CR-gas interaction on much
larger, potentially macroscopic, scales with dramatically reduced computational cost,
while it still retains the full kinetic nature of the CR particles.

Our MHD-PIC code is well suited to investigate a number of astrophysical problems,
as discussed in Section \ref{sec:intro}. In this paper, we have
mainly focused on particle acceleration in non-relativistic shocks, where the Bell
instability plays a dominant role in exciting upstream turbulence and scattering CR
particles. We first point out that when the CR-Hall term is properly taken into account,
the growth rate of the most unstable mode of the Bell instability is reduced
with increasing $\Lambda$,
and it is eventually overtaken by the filamentation instability upon
$\Lambda\gtrsim1$, which explains the findings by \citet{RiquelmeSpitkovsky09}.
We also show that the transition occurs when the most unstable wavelength reaches
the scale of ion inertial length, which is also at the limit of where our
formulation is applicable.

As a proof-of-concept study and non-linear code test, we performed a set of
simulations of parallel MHD shocks with Alfv\'en Mach number $M_A=30$, using an
artificial particle injection prescription with fixed injection efficiency $\eta$.
We observe the development of the Bell instability followed by filamentation and
cavitation as the upstream flow approaches the shock front, leading to strong
magnetic field amplifications and vorticity generation. We find efficient
acceleration of the CR particles, which quickly develop a power-law tail in the
energy spectrum with the expected slope of $f(E)\propto E^{-3/2}$ when all particles
are non-relativistic. The maximum particle energy is found to increase linearly
with time, consistent with efficient particle scattering in the Bohm limit. All
these features are consistent with findings from recent hybrid simulations
\citep{CaprioliSpitkovsky14a,CaprioliSpitkovsky14b,CaprioliSpitkovsky14c},
justifying the ability of our MHD-PIC approach at capturing the non-linear
evolution of CR-driven instabilities and Fermi acceleration in non-relativistic
shocks. We also show that the CR-Hall term can become important in the vicinity
of the shock front in the upstream, especially in the cavities, and for high Mach
number shocks. On the other hand, the CR-Hall effect becomes negligible in the
shock far upstream.

By choosing a reduced particle speed of light ${\mathbb C}$, we further perform
a shock simulation with substantially larger simulation box and longer duration,
and follow particle acceleration from non-relativistic regime to relativistic
regimes. We confirm the expected momentum spectrum of $f(p)\propto p^{-4}$ in
both regimes, with a small drop in normalization
in the relativistic part of the spectrum. We find that maximum particle energy
increases at a slower rate in the relativistic regime (yet still linear in time), which
is likely due to less efficient particle diffusion. In addition, escaping CR particles
are less effective current carriers as they become relativistic, leading to weaker
upstream turbulence and slower particle acceleration.

A major limitation of our initial study of particle acceleration in non-relativistic shocks
is that particle injection is handled artificially. We find that with relatively small
injection efficiency $\eta=10^{-3}$, the shock is less perturbed and particle
acceleration efficiency is low ($\xi\sim6\%$) by the end of the run. The shock
become over-smoothed with large injection efficiency $\eta=4\times10^{-3}$
accompanied by rapid acceleration of particles to $\xi\gtrsim20\%$. Our fiducial choice of
$\eta=2\times10^{-3}$ appears to be optimal and gives acceleration efficiency
$\xi\sim13-15\%$, comparable with self-consistent hybrid-PIC simulations. In reality,
we expect particle injection to be self-adapted to the level such that the shock is not
over-smoothed, maintaining particle acceleration efficiency at some modest level.
In addition, we also expect $\eta$ to gradually decrease with time as particles
are accelerated to higher energies, because the energy content of the power-law
spectrum would diverge with increasing maximum particle energy. Therefore, our
conclusion of optimal injection efficiency $\eta\approx2\times10^{-3}$ mainly
applies to the initial stage of particle acceleration. Nonetheless, for this
effect to be prominent, the CR energy must span several decades in the energy 
spectrum, which is not easily achieved in current simulations.

In the future, we will aim to re-design the particle injection prescriptions
in a way that mimics the particle injection process in hybrid-PIC simulations.
Our current simple prescription fully relies on the shock
detection and tracking algorithm, and its details may have subtle long-term
feedback to the shock structure in the vicinity of the shock front.
Such dependence will be relieved in the new prescriptions. Once properly
calibrated with hybrid-PIC simulations, we will be able to systematically
explore the parameter space of particle acceleration in non-relativistic
shocks, including Alfv\'enic Mach number and magnetic obliquity, and follow
the long-term evolution of the shocks towards larger scales. In particular,
the Alfv\'enic Mach number in supernova remnant shocks is typically several
hundred or higher, which is out of reach of current hybrid-PIC simulations,
but will become feasible with our MHD-PIC approach.

\acknowledgments

We thank the anonymous referee for a very useful report that led to several
improvements to the paper, particularly on code implementation, convergence
checks, and overall clarity.
X.-N.B is supported by NASA through Hubble Fellowship grant
HST-HF2-51301.001-A awarded by the Space Telescope Science Institute, which
is operated by the Association of Universities for Research in Astronomy,
Inc., for NASA, under contract NAS 5-26555.
D.C. is supported by NASA grant NNX14AQ34G.
L.S. is supported by NASA through Einstein Fellowship grant number
PF1-120090 awarded by the Chandra X-ray Center, which is operated by the
Smithsonian Astrophysical Observatory for NASA under contract NAS8-03060. 
This work was partially supported by Simons Foundation (grants 267233  and 291817
to AS), and was facilitated by Max Planck/Princeton Center for Plasma Physics.
Computation for this work was
performed on computational resources supported by PICSciE and the OIT's High
Performance Computing Center at Princeton University. This work also used
resources on Stampede at Texas Advanced Computing Center through XSEDE
grant TG-AST140001.

\appendix

\section[Appendix]{A: Implementation of the MHD-PIC scheme: Details}\label{app:imp}

\subsection[]{``Mass" of CR particles}\label{app:mass}

In our implementation, each CR particle is assigned a ``mass" $m_{\rm CR}$, which
in reality represents the CR mass density carried by the particle (as a swarm of real
particles). The charge and current density carried by particle $j$ in the code is connected
to its ``mass" again via the $q/mc$ factor
\begin{equation}\label{eq:nJCR}
\begin{split}
n_{{\rm CR},j}/c&=(q/mc)_jm_{{\rm CR},j}\ ,\\
{\mb J}_{{\rm CR},j}/c&=(q/mc)_jm_{{\rm CR},j}{\mb u}_j\ .
\end{split}
\end{equation}
The total CR charge and current density $n_{\rm CR}$ and ${\mb J}_{\rm CR}$ at
all grid points can be obtained by particles deposits using the TSC scheme.

We also define the charge-to-mass ratio $(q/mc)_g$ for the thermal ions in the gas,
which is treated to be independent from that for the CR particles. The charge density
of the thermal ions $n_i$ (or $n_g$) is then obtained from the gas density $\rho$ via
\begin{equation}
n_g/c=(q/mc)_g\cdot\rho\ .\label{eq:ng}
\end{equation}
The charge density ratio factor $R$ defined in (\ref{eq:defR}) can then be readily
obtained based on (\ref{eq:nJCR}) and (\ref{eq:ng}).

\subsection[]{The Particle Integrator}

We briefly describe our implementation of the Boris integrator and how it is coupled
with the MHD integrator. Let ${\mb x}_j^{(n)}$ and ${\mb v}_j^{(n)}$ be the location
and velocity of the $j$th particle at step $n$. The integration scheme to step $n+1$
follows the leap-frog pattern in the form of ``drift-kick-rotate-kick-drift":
\begin{equation}
\begin{split}
{\mb x}^{(n+1/2)}_j&={\mb x}^{(n)}_j+\frac{1}{2}{\mb u}^{(n)}_j\Delta t\ ,\\
{\mb v}_j^0&={\mb v}_j^{(n)}+\frac{h}{2}c{\boldsymbol{\mc{E}}}_{j}^{(n+\frac{1}{2})}\ ,\\
{\mb v}_j^1&={\mb v}_j^0+2\frac{{\mb v}_j^0+{\mb v}_j^0\times{\mb b}}
{1+b^2}\times{\mb b}\ ,\\
{\mb v}_j^{(n+1)}&={\mb v}_j^1+\frac{h}{2}c{\boldsymbol{\mc{E}}}_{j}^{(n+\frac{1}{2})}\ ,\\
{\mb x}_j^{(n+1)}&={\mb x}^{(n+1/2)}_j+\frac{1}{2}{\mb u}_j^{(n+1)}\Delta t\ ,\\
\end{split}
\end{equation}
where $h\equiv (q/m_jc)\Delta t$, ${\mb b}\equiv h{\mb B}_j^{(n+1/2)}/\gamma_j$,
and $\gamma_j$ is the Lorentz factor for ${\mb v}_j^0$ and ${\mb v}_j^1$ (which
are the same). Note the relation between ${\mb u}_j$ and ${\mb v}_j$ defined in
(\ref{eq:4vel}).

Electromagnetic fields are extracted from the half time step in the MHD integrator
(with electric field modified by the CR-Hall term), which we denote as
${\boldsymbol{\mc{E}}}^{(n+1/2)}$ and ${\mb B}^{(n+1/2)}$. In addition, while
cell-centered field satisfies ${\boldsymbol{\mc{E}}}\cdot{\mb B}=0$, this is not guaranteed for the
interpolated fields, which may cause spurious particle acceleration. Therefore, we
always clean the parallel component of the interpolated electric field
${\boldsymbol{\mc{E}}}\rightarrow{\boldsymbol{\mc{E}}}-{\boldsymbol{\mc{E}}}\cdot{\mb B}/B$ before it is applied to push the
particles.

The use of mid-point electromagnetic field (which is fixed during the integration)
and the symmetric integration scheme guarantees second-order accuracy in time.
The Boris integrator is time-reversible, and it preserves the geometric properties
of particle gyration in the absence of electric field. Although being straightforward,
we do not implement the Vay's integrator \citep{Vay08}. While it better conserves
particle energy in the ultra-relativistic regime and when the electric field strength
approaches the magnetic field strength, it is computationally more expensive.
Moreover, in non-relativistic MHD, we always have $|c{\mc E}|\ll |{\mathbb C}B|$
(by construction via the choice of ${\mathbb C}$). In this regime, the Vay pusher
behaves essentially the same as the Boris pusher.

\subsection[]{The MHD-PIC Integration Scheme}

We now describe the way to combine the particle integrator with the MHD
integrator in Athena that achieves second-order accuracy in the composite system.
For our purpose, the second-order CTU MHD integrator in Athena can be
considered as a simple predictor-corrector scheme. Our MHD-PIC
scheme is built on top of the existing predictor-corrector framework.
To simplify the notation, we use $f$ and $g$ to represent particle and gas quantities
respectively, and their evolution satisfies the differential equations
\begin{equation}
\frac{df}{dt}=F(f,g)\ ,\quad\frac{dg}{dt}=G(f,g)\ .
\end{equation}
The former describes particle equation of motion under the electromagnetic field
provided by the gas, and the latter represents MHD equations with CR feedback.
Using this notation, our integration scheme can be expressed as follows
\begin{subequations}\label{eq:scheme}
\begin{align}
g^{(n+1/2)}&=g^{(n)}+G(f^{(n)}, g^{(n)})\frac{\Delta t}{2}\ ,\label{seq:gpred}\\
f^{(n+1)}&=f^{(n)}+F(f^{(n+1/2)}, g^{(n+1/2)})\Delta t\ ,\label{seq:boris}\\
g^{(n+1)}&=g^{(n)}+G(\frac{f^{(n)}+f^{(n+1)}}{2}, g^{(n+1/2)})\Delta t\ .\label{seq:corr}
\end{align}
\end{subequations}
In the above, (\ref{seq:boris}) represents the particle integrator described in
the previous subsection. Below, we mainly focus on (\ref{seq:gpred}) and
(\ref{seq:corr}), which represent the predictor and corrector steps for the gas
evolution with CR feedback.

The basic procedure to treat backreaction from CRs to the gas involves adding
source terms to the gas equations by depositing relevant physical quantities
(momentum, energy, etc.) from the location of individual particles to neighboring
grid cell centers.
In the predictor step (\ref{seq:gpred}), the CR feedback is computed using the first
equality of Equations (\ref{eq:gasmotion}) and (\ref{eq:engeq}): we deposit the charge
and current densities of individual particles from their initial positions ${\mb x}_j^{(n)}$
to the grid to obtain $n_{\rm CR}^{(n)}$ and $J_{\rm CR}^{(n)}$, as well as $R^{(n)}$
based on the gas density $\rho^{(n)}$. Together with MHD quantities ${\boldsymbol{\mc{E}}}_0^{(n)}$
and ${\mb B}^{(n)}$, the momentum and energy feedback are directly calculated at cell
centers.

In the corrector step, we take advantage of the fact that all particles have evolved for
a full time step, and compute the momentum and energy difference for each particle
$j$ over one time step:
\begin{equation}
\begin{split}
d{\mb M}_j&=m_{{\rm CR},j}({\mb v}_j^{(n+1)}-{\mb v}_j^{(n)})\ .\\
dE_{k,j}&=m_{{\rm CR},j}(E_{k,j}^{(n+1)}-E_{k,j}^{(n)})\ .
\end{split}
\end{equation}
The opposite of these quantities are deposited to the gas, which account for
momentum and energy feedback from CRs. The deposit is exerted from
particle location at half step ${\mb x}_j^{(n+1/2)}$. This guarantees the
consistency of our algorithm since the feedback is exerted at the same location
as where the particle experiences the Lorentz force. With Athena being a
Godunov code, exact conservation of total momentum and total energy of the
composite gas-CR system is achieved. Moreover, this procedure also ensures
our MHD-PIC algorithm to be second-order accurate.

Regarding the electromagnetic feedback due to the CR-Hall term, Equation
(\ref{eq:emfcr2}) indicates that the correction to the electric field is directly
proportional to the momentum feedback. Therefore, no additional particle deposits
are needed and we simply correct the electric field and the energy flux due to the
CR-Hall term based on the momentum feedback in both the predictor and
corrector steps. The corrections are implemented in accordance with the intrinsic
algorithm that the CTU integrator handles the induction and energy equations.

\subsection[]{Timestepping and Code Performance}\label{app:dt}

The timestepping for the MHD part of the code obeys the standard
Courant-Friedrichs-Lewy condition. Additional constraints on the timestepping from
particles include the following.

First, particle Larmor time $1/\Omega_L=\gamma(qB/mc)^{-1}$ must be resolved.
In practice, we define a number $\Upsilon$ so that we demand
$\Omega_{L, i}\Delta t\leq\Upsilon$ for all particles. Our tests in Section
\ref{sssec:gyration} suggest that $\Upsilon\lesssim0.5$ is generally sufficient. 
For all shock simulations in this paper, we use $\Upsilon=0.3$, and we have also
performed test simulations with $\Upsilon=0.1$ and $0.5$ and do not find any practical
difference in the outcomes.

Second, individual particles can not travel more than two grid cells per time step.
This follows from the specifics of our MHD-PIC scheme\footnote{With the particle
module turned on in Athena, the innermost two (otherwise just one) ghost zones
are updated during the predictor step. The TSC scheme requires that particles can
move to the innermost but not the next ghost zone in the predictor step.}.

Since CR particles can become relativistic, with velocities much larger than typical
fluid velocity, the code can be slowed down substantially. To reconcile this issue, we
have implemented sub-cycling for the CR particles: we evolve multiple steps of particles
per MHD step. In other words, the single-step particle evolution illustrated in
(\ref{seq:boris}) is now replaced by multiple steps of particle updates. When evolving
the particles, we always use the electromagnetic field at half MHD step (i.e.,
$g^{(n+1/2)}$), hence the MHD-PIC scheme remains second-order accurate over a
full MHD step. Particles deposit feedback at each sub-step, and the feedback
accumulated from all sub-steps will finally be added to the gas in the corrector step as usual.

We have tested the code performance on a single CPU as well as in parallel. For
particles, most of the computation time is spent on the Boris integrator, particle
feedback and various interpolations. The serial test takes place on a Intel Xeon E5
2.3GHz processor. Without sub-cycling, it takes about $1\mu$s to integrate one particle
in 3D, and $0.6\mu$s to integrate one particle in 2D. Combined with isothermal MHD
using the CTU integrator with third-order reconstruction and the HLLD solver, we find
that the code spends roughly equal amount of time on the MHD integrator and on the
particles when using approximately 7 (for 3D) or 6 (for 2D) particles per cell.
When run in parallel, because of extra steps for feedback exchange, the particle
numbers quoted above are reduced to 6 (for 3D) and 5 (for 2D).
The code performance on the particles can be enhanced by using sub-cycling.
With the same test on the single processor, we find that the averaged time to
integrate one particle for one particle sub-step is reduced by $30\%$ when using 10
particle sub-steps per MHD step.

\section[]{B: Code Tests}\label{app:test}

The two main ingredients of our MHD-PIC code are the Boris integrator and the particle
feedback. We outline here two problems that test the two ingredients separately,
and show that the code performs well in these respects. Since in Athena the units
for magnetic field are such that magnetic permeability $\mu=1$, factors of $4\pi$
will be left out here as well.

\begin{figure*}
    \centering
    \includegraphics[width=160mm]{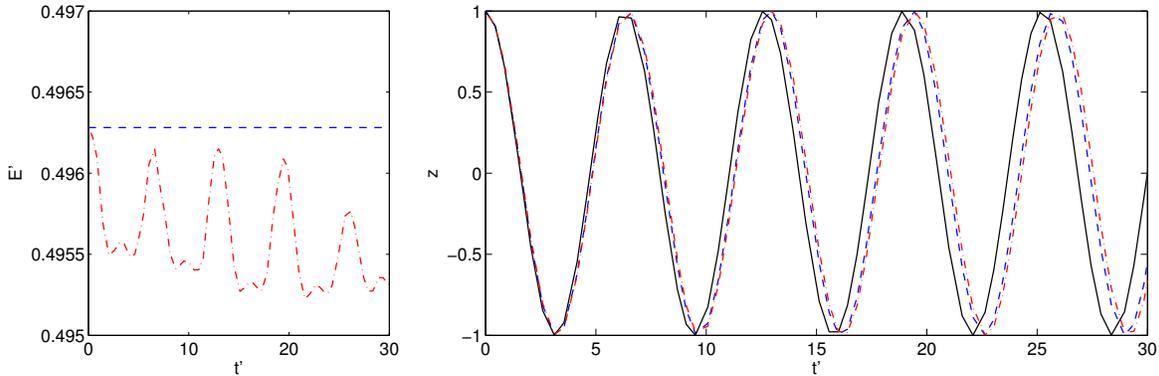}
  \caption{Gyration test for non-relativistic particle orbits with $v_{\perp}=0.1{\mathbb C}=1$
  using variable time step with $\Omega_L\Delta t=0.5\pm0.1$.
  Left: time evolution of particle kinetic energy in the co-moving frame (indicated by the prime).
  Right: evolution of particle position in the co-moving frame.
  Blue dashed lines correspond to zero background flow velocity ($v_0=0$), red
  dash-dotted lines correspond to mildly relativistic drift of the background gas
  ($v_0=1$). Black solid lines represent analytical results. Note that we
  have chosen to a large time step to exaggerate the truncation errors.}\label{fig:Boris_NR}
\end{figure*}

\begin{figure*}
    \centering
    \includegraphics[width=160mm]{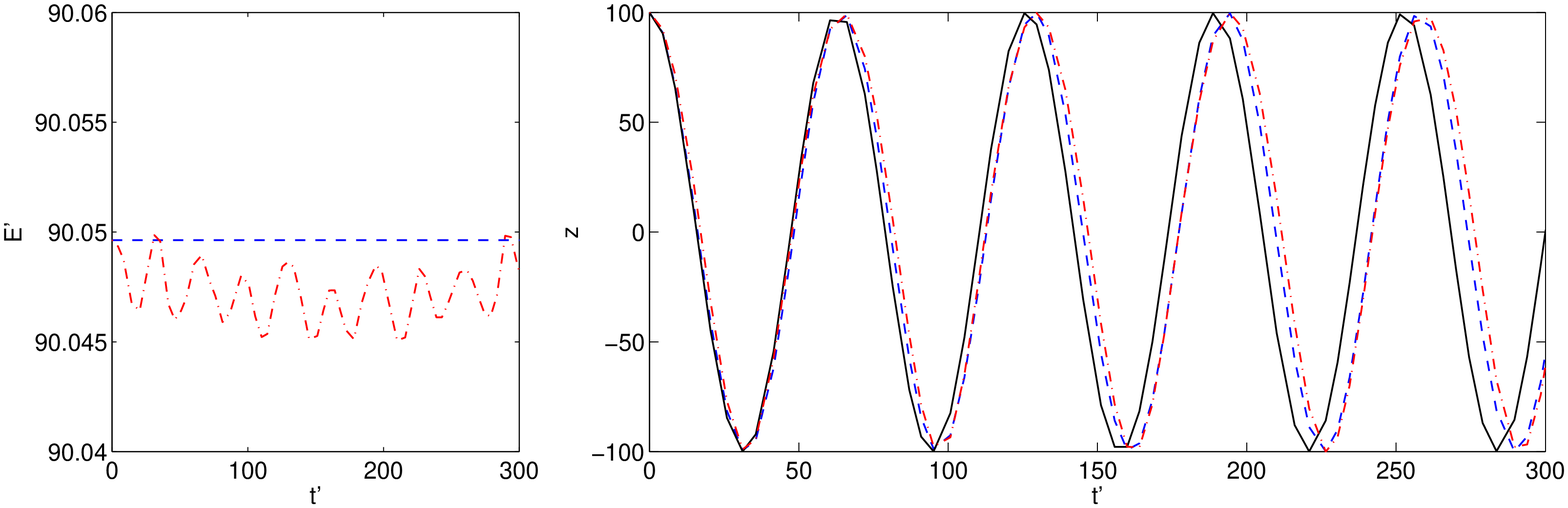}
  \caption{Same as Figure \ref{fig:Boris_NR}, but for relativistic particles with
  $v_\perp=10{\mathbb C}$.}\label{fig:Boris_SR}
\end{figure*}

\subsection[]{Gyration Test}\label{sssec:gyration}

We test the Boris integrator with the particle gyration problem. We set up a uniform
background gas that moves at constant velocity $v_0$ in the $\hat{y}$ direction.
The velocity can be ``relativistic" based on the artificial speed of light ${\mathbb C}$,
with Lorentz factor $\gamma_0={\mathbb C}/\sqrt{{\mathbb C}^2-v_0^2}$.
To facilitate analytical calculation, we prescribe other parameters in a frame that is
{\it co-moving} with the gas. In the co-moving frame, the electric field is zero, and there
is a uniform magnetic field $B_0$ along the $\hat{x}$ direction. We include one
particle whose parallel and perpendicular four-velocities in the co-moving frame are
given by $v_{\parallel}$ and $v_{\perp}$, with Lorentz factor
$\gamma_1=\sqrt{{\mathbb C}^2+v_{\parallel}^2+v_\perp^2}/{\mathbb C}$. Then we
have the Larmor radius $R_L=v_{\perp}(qB_0/mc)^{-1}$, gyro-frequency
$\Omega_L=(qB_0/\gamma_1mc)$, parallel velocity $u_0=v_{\parallel}/\gamma_1$.
We compute the particle orbit in the lab frame using the code and compare the results
with analytical orbits in the co-moving frame.


In Figure \ref{fig:Boris_NR}, we show the test results for non-relativistic particles. We
initialize the problem with $B_0=q/mc=v_{\perp}=1$, $v_{\parallel}=0$, ${\mathbb C}=10$,
and hence $R_L=1$ and $\Omega_L=1$. Considering the fact that the time step in our
MHD-PIC code is generally not a constant, the time step we adopt for this test is
$\Delta t=0.5+0.1\cos\alpha$, where $\alpha$ is a random number. Two tests are performed,
one with zero drift velocity $v_0=0$ (blue dashed), and the other with drift velocity $v_0=1$
(red dash-dotted). Shown on the left is the evolution of particle kinetic energy in the co-moving
frame, while on the right is the particle trajectory in the co-moving frame. We see that for
$v_0=0$, the integrator conserves energy exactly despite the variable time step, with
relatively small phase error. With drift $v_0=1$, energy is no longer conserved. There is
systematic oscillation in $E'$ is at $0.1\%$ level, which is due to non-zero background
electric field.\footnote{The level of energy oscillation increases rapidly with increasing
gas flow velocity $v_0$ once $v_0$ exceeds particle velocity. Nevertheless, for all potential
applications of our code, we expect CR particle velocity to be (much) larger than $v_0$.}
The energy $E'$ also undergoes a random-walk type drift over time,
which is due to the variable time step. Note that both oscillation and drift in $E$ are inherent
to the Boris pusher under these circumstances.

\begin{figure*}
    \centering
    \includegraphics[width=180mm]{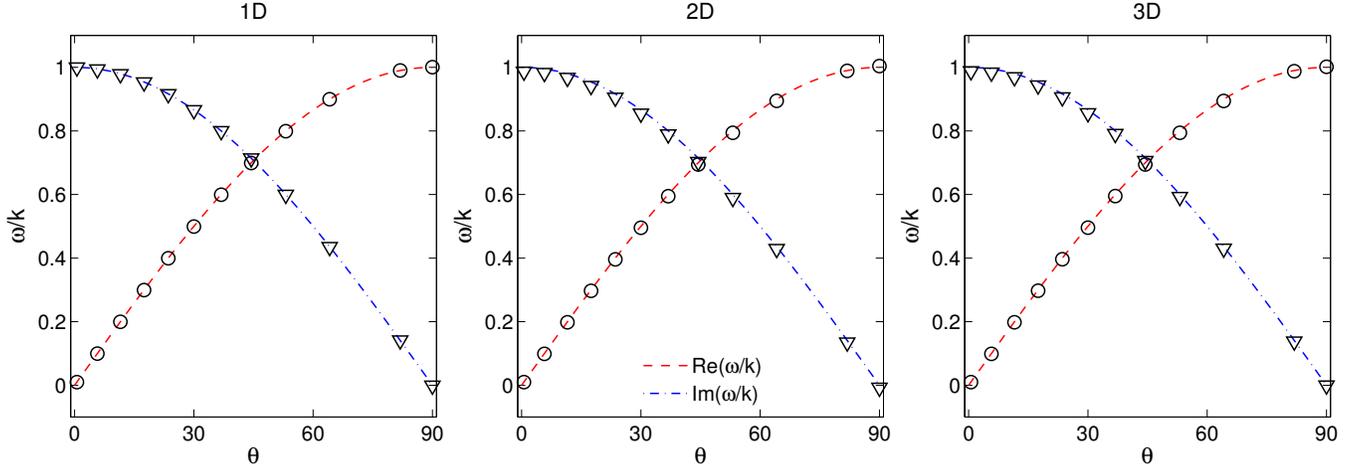}
  \caption{Linear dispersion relation test of the Bell instability in 1D, 2D and 3D with
  $\Lambda\ll1$ (the CR-Hall term is negligible). The real (red dashed and circles) and
  imaginary (blue dashh-dotted and triangles) parts of the most unstable mode as a
  function of $\theta$, defined as $\sin\theta=\epsilon$, where $\epsilon\equiv v_A/u_0$,
  the ratio of Alfv\'en velocity to the CR drift velocity. Symbols denote measured values
  from the code test, curves represent analytical expectations. See Section
  \ref{ssec:bellgrow} for details.}\label{fig:CRDI-disp}
\end{figure*}

\begin{figure*}
    \centering
    \includegraphics[width=180mm]{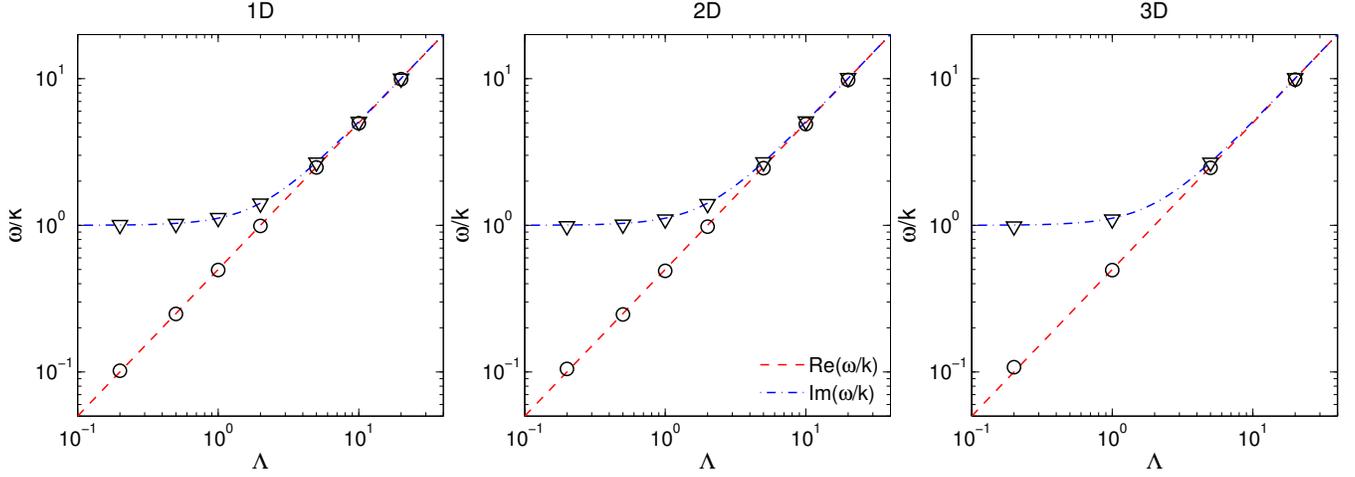}
  \caption{Linear dispersion relation test of the Bell instability in 1D, 2D and 3D with
  $\epsilon\ll1$ (the CR charge density is negligible). The real (red dashed and circles)
  and imaginary (blue dashh-dotted and triangles) parts of the most unstable mode as a
  function of $\Lambda$, defined as $\Lambda=R/\epsilon$.
  Symbols denote measured values
  from the code test, curves represent analytical expectations. See Section
  \ref{ssec:bellgrow} for details.}\label{fig:CRH-disp}
\end{figure*}

In Figure \ref{fig:Boris_SR}, we show the results from the same test but for relativistic particles.
We increase particle velocity to $v_{\perp}=100$ ($\gamma_1\approx10$), keeping other
parameters fixed. Correspondingly, we have $R_L=100$ and $\Omega_L\approx0.1$,
and we modify the time step to $\Delta t=5\pm\cos\alpha$ so that $\Omega_L\Delta t$
remains the same as before. We see again that without $E\times B$ drift, the integrator
captures particle orbits perfectly well. In the presence of drift, the fractional error in the
energy is smaller than in the previous case, while the phase error remains similar.

Note that we have used relatively large time step ($\Omega_L\Delta t=0.5$) to exaggerate
the truncation errors. The time step in real applications is typically much smaller. Moreover,
random variation in the timestep also enhances energy drift in the tests, while in real
simulations, timestep variation is smooth.
Therefore, our particle gyration tests have demonstrated a reliable performance of
the Boris integrator for astrophysical applications.

\subsection[]{Linear Growth of the Bell Instability}\label{ssec:bellgrow}

We test the CR feedback and the CR-Hall term using the linear dispersion
relation of the Bell instability, as discussed in Section \ref{sec:crcdi}. The test is
conducted by evolving the exact eigen-vectors of the most unstable mode.

Two sets of tests are performed. In the first test, we suppress the CR-Hall
term by using sufficiently small $R$ such that $\Lambda\equiv RU_s/v_A\ll1$.
In this case, the most unstable wavenumber is $k_0$, and the dispersion relation
(\ref{eq:belldisps}) gives
\begin{equation}
\omega=k_0v_A(\epsilon+{\rm i}\sqrt{1-\epsilon^2})
=k_0v_A\exp{\bigg[{\rm i}\bigg(\frac{\pi}{2}-\theta\bigg)\bigg]}\ ,\label{eq:disp_mu}
\end{equation}
where we have defined $\sin\theta\equiv\epsilon$. The corresponding eigenvector of
this mode reads
\begin{equation}\label{eq:eigenvect}
\begin{split}
&b_y=b_\perp\cos{(k_0x-\epsilon k_0v_At)}\ ,\\
&b_z=b_\perp\sin{(k_0x-\epsilon k_0v_At)}\ ,\\
&u_y=\frac{v_A}{B_0}b_\perp\sin{(k_0x-\theta-\epsilon k_0v_At)}\ ,\\
&u_z=-\frac{v_A}{B_0}b_\perp\cos{(k_0x-\theta-\epsilon k_0v_At)}\ ,\\
\end{split}
\end{equation}
where ${\mb b}$ and ${\mb u}$ denote transverse perturbations of the magnetic
field and fluid velocity, and the wave amplitude grows as
$b_\perp\propto\exp{(\cos\theta k_0v_At)}$. 

To setup the problem, we consider background gas with uniform density $\rho_0$
and zero velocity. Let ${\mb B}_0$ be the background magnetic field,
${\mb J}_{\rm CR}=n_{\rm CR}{\mb u}_0$ be the external uniform current density provided
by the CR particles, both of which are along the $\hat{x}$ direction. Note that the CR current
is provided by a spatially uniform distribution of CRs with number density $n_{\rm CR}$
that drift at velocity $u_0$. For our code test purpose, we treat the CR particles as having
infinite inertia so that the their effects are simply to provide constant external current and
charge. 

In our code test, we set up the exact eigenvector of the most unstable mode
(\ref{eq:eigenvect}), and test the dispersion relation (\ref{eq:disp_mu}) as a
function of $\epsilon$ (or $\theta$). To elaborate, we consider the following
parameters: $B_0=\rho_0=1$ (hence $v_A=1$). We fit one most unstable
wavelength $\lambda_0$ into the simulation box, with $\lambda_0$ determined by
the box size set by the user. This fixes the CR current density:
$J_{\rm CR}=2B_0k_0=4\pi/\lambda_0$. We initiate with a cold population of
particles moving at constant velocity $u_0=v_A/\epsilon$ along the direction of
${\mb B}_0$. There is one particle per grid cell located cell centers. The main
parameter is $\epsilon$ ranging from 0 to 1. Correspondingly, the growth rate of
the instability is simply $\sqrt{1-\epsilon^2}k_0$.
In order for the Lorentz force to be negligible on particle motion over a few
$e$-folding time, we set $q/mc\equiv10^{-6}k_0$ for the CR particles. This further
determines the value of $m_{\rm CR}$, which is simply given by
$m_{\rm CR}=J_{\rm CR}/[(q/mc)u_0]=2\times10^{6}\epsilon$.
We finally set the gas isothermal sound speed to be $c_s=1$, and speed of light
${\mathbb C}=100u_0$, both of which are irrelevant for this problem but needed in the code.

The problem generator deals with all 1D, 2D and 3D cases, where in multi-dimensions,
the wave vector of the mode is not grid-aligned. The box size in these cases are $1$,
$\sqrt{5}\times\sqrt{5}/2$, $3\times1.5\times1.5$, resolved by $32$, $64\times32$ and
$96\times48\times48$ cells respectively. The choice of such box size
and resolution guarantees that $\lambda_0=1$ with approximately $32$ cells per
wavelength. We vary $\epsilon$ from $0.01$ to $1.0$, and show the measured real and
imaginary parts of $\omega/k$ in Figure \ref{fig:CRDI-disp} as a function of $\theta$
where $\sin\theta=\epsilon$. Theoretically, the real and imaginary parts should
be exact sine and cosine curves, and we see that in all three dimensions the code very
well reproduces the analytical phase velocity and growth rate. Although not obvious in
the plot, deviations from analytical results in 2D and 3D cases are comparable and
larger than the 1D case, but are well within $2\%$.

In the second test, we relax the constraint on $\Lambda$ while keeping $R\ll1$, which
enables the CR-Hall term without violating our formulation. We also keep
$\epsilon\ll1$. In this case, the most unstable wavenumber is $k_m=k_0/f$, where
$f=1+(\Lambda/2)^2$, and the dispersion relation reads
\begin{equation}
\omega=(k_mv_A)(\Lambda/2+{\rm i}{\sqrt{f}})\ .
\end{equation}
The corresponding eigen-vectors are
\begin{equation}\label{eq:eigenvect2}
\begin{split}
&b_y=b_\perp\cos{\phi}\ ,\\
&b_z=b_\perp\sin{\phi}\ ,\\
&u_y=\frac{v_A}{B_0}b_\perp[(\Lambda/2)\cos\phi+\sqrt{f}\sin{\phi}]\ ,\\
&u_z=\frac{v_A}{B_0}b_\perp[(\Lambda/2)\sin\phi-\sqrt{f}\cos{\phi}]\ ,\\
\end{split}
\end{equation}
where $\phi\equiv k_m[x-(\Lambda/2)v_At]$, and the wave amplitude grows as
$b_\perp\propto\exp(k_mv_A\sqrt{f}t)$.

We use the same numerical setup as the previous test, except for resetting the
eigenvectors. By varying $\Lambda$ from 0.2 to 20, we show the measured real
and imaginary parts of $\omega/k$ in Figure \ref{fig:CRH-disp}. We see that for
all values of $\Lambda$, the phase velocity and growth rate are accurately
reproduced from the regime $\Lambda\ll1$ to $\Lambda\gg1$. In particular, the
measured growth rates deviate from analytical results by less than $0.5\%$ for
all cases at our numerical resolution. Although our code is may not be physically
reliable if $\Lambda\gtrsim1$ (see Section \ref{ssec:mhdcr}), this test demonstrates
the successful  implementation of the CR-Hall term.

\section[]{C: Linear Dispersion Relation of the Bell Instability}\label{app:bell}

In this Appendix, we provide more detailed derivation of the linear dispersion
relation of the Bell instability taking into account of the CR-Hall term, following
Equation (\ref{eq:bell1}) in Section \ref{sec:crcdi}. We first define
\begin{equation}\label{eq:transform}
\tilde{\omega}\equiv\omega-(\Lambda/2)v_Ak\ ,\qquad
\tilde{U}_s\equiv U_s(1-R/2)\ ,
\end{equation}
and two factors
\begin{equation}
g\equiv(1-R)(1-R/2)\ ,\qquad f\equiv1-R+(\Lambda/2)^2\ .
\end{equation}
Now Equation (\ref{eq:bell1}) can be rewritten with the new variables
\begin{equation}
{\rm i}(\tilde{\omega}^2-fk^2v_A^2)({\mb b}\times\hat{x})
+g\frac{kJ_{\rm CR}v_A^2}{B_0}
\bigg(1-\frac{\tilde{\omega}}{k\tilde{U}_s}\bigg){\mb b}=0\ .
\end{equation}
Non-trivial solution requires $b_y=\pm{\rm i}b_z$, corresponding to left/right
circularly polarized modes, and it is straightforward to obtain the dispersion relation
\begin{equation}
\tilde{\omega}^2-fk^2v_A^2=\pm g\frac{kJ_{\rm CR}v_A^2}{B_0}
\bigg(1-\frac{\tilde{\omega}}{k\tilde{U}_s}\bigg)\ .\label{eq:belldisp1}
\end{equation}
To make it more instructive, we further define
\begin{equation}
k_0\equiv\frac{J_{\rm CR}}{2B_0}\ ,\qquad
\epsilon\equiv\frac{v_A}{\tilde{U}_s}\ ,
\end{equation}
and the above dispersion relation can be rewritten as
\begin{equation}
(\tilde{\omega}\pm\epsilon k_0v_A)^2
=f\bigg(k\pm \frac{g}{f}k_0\bigg)^2v_A^2-(k_0v_A)^2\bigg(\frac{g^2}{f}-\epsilon^2\bigg)\ ,\label{eq:belldisp2}
\end{equation}
where the two $\pm$ signs must be taken to be the same. It reduces to Equation (\ref{eq:belldisps})
when $R\ll1$, or $g=1$.

\bibliographystyle{apj}

\label{lastpage}
\end{document}